\documentclass{sig-alternate} 
\usepackage{mathptmx} % This is Times font

\newcommand{\ignore}[1]{}
\usepackage{fancyhdr}
\usepackage[normalem]{ulem}
\usepackage[hyphens]{url}
\usepackage{microtype}

\usepackage{amsmath}                        % better math environment
\usepackage{epigraph}
\usepackage[all]{xy}
\usepackage[hang]{subfigure}                % allow subfigures
\usepackage{url}                            % typeset urls
\usepackage{amssymb, amsmath, subfigure, acronym}
\usepackage{multirow}
\usepackage{array}
\usepackage{tabularx}
\usepackage{graphicx}
\usepackage{cite}
\usepackage{epstopdf}
\usepackage{csquotes}

\makeatletter
\DeclareRobustCommand{\rvdots}{%
  \vbox{
    \baselineskip4\p@\lineskiplimit\z@
    \kern-\p@
    \hbox{.}\hbox{.}\hbox{.}
  }}
\makeatother

\newcommand\MyLBrace[2]{%
\left.\rule{0pt}{#1}\right\}\text{#2}}

\usepackage{arydshln}

\newcommand{\fixme}[2]{\ifx&#2&{\leavevmode\color{red}#1}\else{\leavevmode\color{red}FIXME\{}#1{\leavevmode\color{red}\}}\footnote{{\leavevmode\color{red}#2}}\PackageWarning{Fixme}{#1: #2}\fi}

\usepackage[usenames, dvipsnames]{color}
\usepackage{arydshln}
\usepackage{bm}

\usepackage{tikz}
\usetikzlibrary{arrows,%
                petri,%
                topaths,
                shapes}%
\usetikzlibrary{shapes.geometric}
\usepackage{siunitx}
\usepackage{tkz-berge}
\usepackage{pgfplots}

\pgfplotsset{width=7cm,compat=1.3}
\usepackage{booktabs}
\usepackage{multicol}

\usepackage{tablefootnote}

\usepackage[para]{threeparttable}

\usepackage[linesnumbered,ruled]{algorithm2e}

\usepackage[bookmarks=true,breaklinks=true,letterpaper=true,colorlinks,linkcolor=black,citecolor=blue,urlcolor=black]{hyperref}

\newcommand{\iscasubmissionnumber}{478}
\fancypagestyle{firstpage}{
  \fancyhf{}
\renewcommand{\headrulewidth}{0pt}
  \fancyhead[C]{\normalsize{ISCA 2018 Submission
      \textbf{\#\iscasubmissionnumber} \\ Confidential Draft: DO NOT DISTRIBUTE}} 
  \fancyfoot[C]{\thepage}
}  

\title{Multi-Mode Inference Engine for Convolutional Neural Networks \vspace{-20pt} } 

\author{Arash~Ardakani, Carlo~Condo and Warren~J.~Gross\\
Electrical and Computer Engineering Department, McGill University, Montreal, Quebec, Canada}

\begin{document}
\maketitle

\pagestyle{plain}

\begin{abstract}
During the past few years, interest in convolutional neural networks (CNNs) has risen constantly, thanks to their excellent performance on a wide range of recognition and classification tasks. However, they suffer from the high level of complexity imposed by the high-dimensional convolutions in convolutional layers. Within scenarios with limited hardware resources and tight power and latency constraints, the high computational complexity of CNNs makes them difficult to be exploited. %As a result, plenty of software and hardware solutions have been introduced in literature to reduce the complexity of CNNs. On the software side, the research on CNNs has mainly focused on reducing the number of operations of CNNs.
%On the other hand, 
Hardware solutions have striven to reduce the power consumption using low-power techniques, and to limit the processing time by increasing the number of processing elements (PEs). While most of ASIC designs claim a peak performance of a few hundred giga operations per seconds, their average performance is substantially lower when applied to state-of-the-art CNNs such as AlexNet, VGGNet and ResNet, leading to low resource utilization. Their performance efficiency is limited to less than $55\%$ on average, which leads to unnecessarily high processing latency and silicon area. In this paper, we propose a dataflow which enables to perform both the fully-connected and convolutional computations for any filter/layer size using the same PEs. We then introduce a multi-mode inference engine (MMIE) based on the proposed dataflow. Finally, we show that the proposed MMIE achieves a performance efficiency of more than 84\% when performing the computations of the three renown CNNs (i.e., AlexNet, VGGNet and ResNet), outperforming the best architecture in the state-of-the-art in terms of energy consumption, processing latency and silicon area.
\end{abstract}

\section{Introduction} \label{intro}
Deep neural networks (DNNs), especially convolutional neural networks (CNNs) \cite{lenet}, have received tremendous attention due to their ability to surpass human-level accuracy on a wide range of complex tasks such as recognition, classification and detection \cite{deeplearning}. Depending on their size and complexity, these networks achieve different degrees of classification/recognition accuracy. A CNN is a stack of multiple convolutional layers followed by fully-connected layers: they extract high level abstractions and features of raw data, whereas fully-connected networks are used to learn non-linear combinations of the extracted features. In 2012, a CNN called AlexNet \cite{alexnet} was introduced: it is constituted of 5 convolutional layers followed by 3 fully-connected layers and achieves $42.9\%$ misclassification rate (MCR) on the ImageNet dataset. AlexNet contains 2.3M weights and 58.6M weights in its convolutional and fully-connected layers, respectively, performing 1332M operations (i.e., 666M multiplications-accumulations) in its convolutional layers and 117.2M operations (i.e., 58.6M multiplications-accumulations) in its fully-connected layers. VGGNet-16 \cite{vgg} is another well-known CNN, containing 13 convolutional layers with 14.7M weights and 3 fully-connected layers with 124M weights. VGGNet-16 performs 30.6G operations in its convolutional layers and 248M operations in its fully-connected layers, achieving $27\%$ MCR on ImageNet. Recently, ResNet-50 \cite{resnet}, containing 49 convolutional layers with 23.5M weights and 1 fully-connected layer with 2M weights, achieved a better MCR (i.e., 22.85\% on ImageNet) by going even deeper. ResNet-50 respectively performs 7G and 4M operations within the two types of layers. All these CNNs have won the ImageNet Large Scale Visual Recognition Challenge (ILSVRC) \cite{ILSVRC}.\par

Regardless of the fact that in almost all the aforementioned CNNs the majority of weights is found in fully-connected layers, the number of operations are dominated by convolutions. As a result, the processing time of CNNs is also dominated by the convolutional processes. This issue can easily be addressed by exploiting parallel processing elements (PEs) to increase throughput. However, a straightforward parallelization requires high data movement and bandwidth, leading to high energy consumption \cite{conv_imp2}. It is worth noting that memory accesses to off-chip memories are more expensive than on-chip storage, as shown in \cite{energy}.\par

%\fixme{As a first attempt to address the above issues, spiking neural networks and stochastic models were introduced in \cite{truenorth, ISC,sips,ISC1,sbnn,lsdl}, where numbers are represented as sequences of random bits, allowing to perform computations using simple logic gates. Although the simplicity of stochastic computations leads to very low-power implementations, the inherent long latency of these paradigms makes them unfit for real-time applications.}{Can probably go if in need of space.}

Pruning techniques were first introduced in \cite{pruning, pruning1} to reduce the number of parameters and memory accesses to off-chip memory. In \cite{pruning} CPU/GPU implementations were considered, showing that 3$\times$ to 4$\times$ layer-wise speedup can be obtained for fully-connected layers without any practical speedup for convolutional layers. To accelerate convolutional processes on GPUs and CPUs, a new method was also introduced in \cite{ssl}, achieving up to 5.1$\times$ speedup. The work presented in \cite{isca} introduces a fully-connected accelerator, called efficient inference engine (EIE), for the pruning technique introduced in \cite{pruning, pruning1}. EIE can obtain 13$\times$ to 307$\times$ speedup, and save 2700$\times$ to 24000$\times$ energy compared to CPUs or GPUs for fully-connected computations. Recently, a new pruning technique and its custom hardware were introduced in \cite{sfc}, using low-cost linear-feedback shift registers (LFSRs) to prune the connectivity of fully-connected layers. This technique also saves up to 90\% energy compared to conventional implementations of fully-connected layers. However, as discussed earlier, convolutional processes are the bottleneck of the processing time of CNNs.\par

During the past few years, many convolutional accelerators with different dataflows have been introduced in literature \cite{eyeriss1,moons_new,DaDianNao,origami,1dconv,DNPU}. While these ASIC architectures can successfully reduce the energy consumption of convolutional processes and meet the latency constraints of small CNNs such as AlexNet, they fail to employ the full potential of their architectures, resulting in a low performance efficiency. In fact, there is a huge gap between their peak performance and average runtime performance. 
% The peak performance is defined as the maximum achievable number of operations per second, roughly computed as
% \begin{equation}
% \text{peak performance} = 2 \times \#\text{PE} \times f,
% \end{equation}
% \fixme{where $\#$PE and $f$ denote the number of PEs and nominal frequency, respectively, and where each PE performs multiplications-accumulations (MACs). The runtime performance is also defined as the number of operations per second obtained when performing computations on a specific dataset. Therefore, the performance efficiency is computed as the ratio between the average runtime performance and the peak performance.}{All this is interesting but long. If we need space, we might try to shrink it.} 
For instance, in \cite{eyeriss1} the architecture known as Eyeriss achieves a peak performance of 84 Gops, where each MAC is considered as two operations. However, its performance efficiency is limited to 55\% and 26\% when performing the convolutional computations of AlexNet and VGGNet-16, respectively.\par

To improve the performance efficiency and to accelerate the convolutional processes for VGG and VGG-like networks, a dataflow, called fully-connected inspired dataflow (FID), and the architecture implementing it were introduced in \cite{TCAS}. This architecture achieves a high performance efficiency of $90\%$ on the convolutional processes of VGGNet-16. Despite its high performance efficiency, throughput and low silicon area, it is only limited to architectures with $3 \times 3$ filters.\par

In this paper, we propose a dataflow supporting all type of filter sizes used in state-of-the-art CNNs by generalizing FID. We provide a theoretical framework showing that the proposed generalized FID (GFID) can perform both the fully-connected and convolutional processes while using the same hardware resources, resulting in a high utilization factor. We then propose a CNN accelerator based on the proposed GFID, that performs both fully-connected and convolutional computations, which is hereafter referred to as multi-mode inference engine (MMIE). MMIE is optimized to achieve high performance efficiency and low memory accesses to the off-chip memory, while keeping the power consumption below the budget of mobile/embedded devices. Finally, we evaluate the performance of MMIE on the state-of-the-art CNN models (i.e., AlexNet, VGGNet-16 and ResNet-50) and show that MMIE performs the convolutional computations of these CNNs with an $83\%$ minimum performance efficiency.

\section{Preliminaries} \label{pre}
%\subsection{Fully-Connected Networks}\label{pre:fc}
A fully-connected network is a stack of layers where each neuron is connected to every neuron in the previous and next, and to each connection is associated a weight $w$. A fully-connected layer performs the following computations with $n$ inputs and $m$ outputs:
\begin{equation}
y = \text{ReLU}(w_{m\times n}x_{n\times 1} + b_{m\times 1}),
\label{fc_comp}
\end{equation}
where $x$ denotes the input pixels, $y$ the output pixels, $b$ the biases, and $\text{ReLU}$ is the non-linear activation function $\text{ReLU} = \max(0,x)$. According to (\ref{fc_comp}), the fully-connected computational kernel calculates numerous vector-matrix multiplications followed by the $\text{ReLU}$. Due to parallel memory access requirement for fully-parallel implementations of such networks, a semi-parallel implementation is a typical approach for their hardware implementations \cite{TCAS}. In semi-parallel implementations, only a limited number of PEs is instantiated, and computations for each neuron are performed serially \cite{conv_neuron}. In fact, different trade-offs between area occupation and latency can be obtained by changing the degree of parallelism.\par

%\subsection{Convolutional Layers}\label{pre:conv}
 Inspired by the organization of the animal visual cortex, it was shown that the connectivity of neurons in convolutional layers can be mathematically described by a convolution operation \cite{conv}. All neurons in a convolutional layer share a set of weights, also referred to as a filter.\par

\begin{figure}[!t]
\centering
\includegraphics[scale = 0.5]{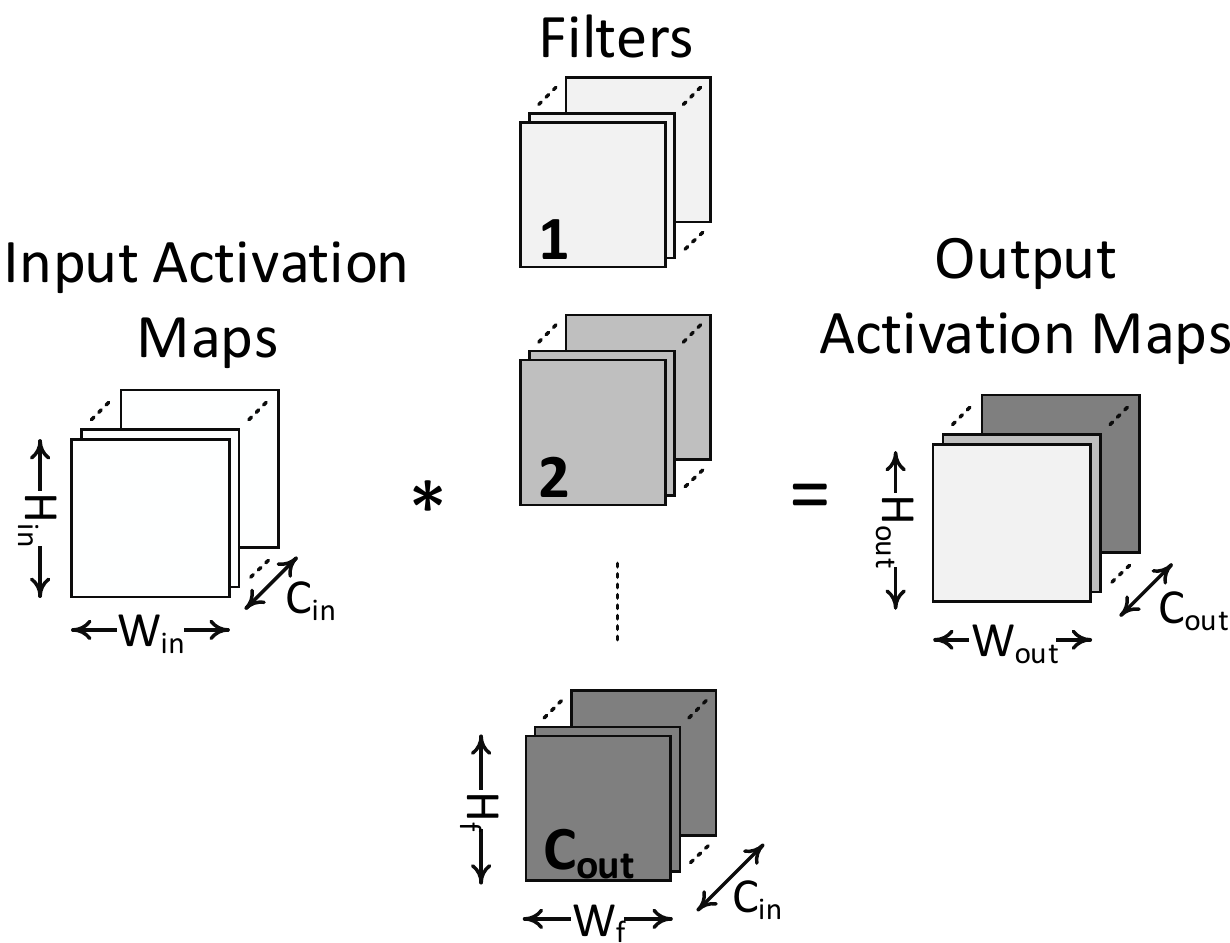}
\caption{The high-dimensional convolutions in a convolutional layer.}
\label{CL}
\end{figure}

The main computational kernel of a convolutional layer involves high-dimensional convolutions, as shown in Fig. \ref{CL}. The convolutional layers take input pixels, which are also called input activation maps, arranged in 3 dimensions (i.e., height $H_{in}$, width $W_{in}$ and channel $C_{in}$), and generate output pixels, which are also called output activation maps, arranged in 3 dimensions (i.e., height $H_{out}$, width $W_{out}$ and channel $C_{out}$). This transformation is a result of the convolution between the input activation maps and a set of $C_{out}$ 3D filters. More precisely, every single 2D $H_{out} \times W_{out}$ plane of the output activation maps is a result of the convolution between the 3D input activation maps with a set of 3D filters. In fact, a summation of multiple plane-wise 2D convolutions forms a 3D convolution. At the end, the results of 3D convolutions are also added to 1D bias. In summary, the convolutional processes with the input activation maps, the output activation maps, the filters and the bias matrices denoted as $X$, $Y$, $W$ and $B$, respectively, can be expressed as
\begin{equation}
\small
Y(z,t,q) = B(q) + \sum_{k = 1}^{C_{in}} \sum_{j = 1}^{H_f} \sum_{i = 1}^{W_f} X(zS + j,tS + i,k) \times W(j,i,k,q), \nonumber
\end{equation}
\begin{equation}
H_{out} = (H_{in} - H_{f} + S)/S, \nonumber
\end{equation}
\begin{equation}
W_{out} = (W_{in} - W_{f} + S)/S,
\label{eq_conv}
\end{equation}
where $1 \leq z \leq H_{out}$, $1 \leq t \leq W_{out}$ and $1 \leq q \leq C_{out}$. The stride $S$ represents the number of activation map pixels of which the filter is shifted after each convolution. %Table \ref{conv_par} summarizes the aforementioned parameters. 
Contrary to the fully-connected layers, convolutional computations are dominated by numerous MACs according to Eq. (\ref{eq_conv}), leading to a high degree of computational complexity.
% 
% \begin{table}[!t]
% \renewcommand{\arraystretch}{1.3}
% \renewcommand{\thefootnote}{\alph{footnote}}
% \caption{The Required Parameters for Convolutional Processes. \fixme{Avoidable in case of need for extra space.}{}}
% \centering
% 
% \scalebox{0.96}{
% \begin{tabular}{c c}
% \hline
% \hline
% \multicolumn{1}{c}{Name of Parameter} & \multicolumn{1}{c}{Description}\\
% \hline
% $H_{in}$/$W_{in}$ & Height/Width of the input activation maps \\
% \hline
% $H_{out}$/$W_{out}$ & Height/Width of the output activation maps \\
% \hline
% $H_{f}$/$W_{f}$ & Height/Width of the filter plane\\
% \hline
% $C_{in}/C_{out}$ & No. of channels of the input/output activation maps \\
% \hline
% $S$ & Stride number \\
% \hline
% \hline
% \end{tabular}}
% \label{conv_par}
% \end{table}

\subsection{Fully-Connected Inspired Dataflow for Convolutional Computations}\label{pre:FID}
In \cite{TCAS}, FID was introduced. It can be used to efficiently perform the computations of convolutional layers with filter parameter $W_f$ fixed to $3$. Let us note that 2D convolution is the weighted summation of each pixel of an input image with its neighboring pixels, and consider an input image as a matrix $X_{8\times 2}$, a filter as a matrix $W_{1\times 3}$ and an output as a matrix $Y_{6\times 2}$, such that
\[
X =
\begin{bmatrix}
    X_{1} & X_{2} & \dots  & X_{8} \\
    X_{9} & X_{10} & \dots  & X_{16} \\
\end{bmatrix}
, W =
\begin{bmatrix}
    W_{1} & W_{2} & W_{3} \\
\end{bmatrix},
\]\vspace{-10pt}
\[
Y =
\begin{bmatrix}
    Y_{1} & Y_{2} & \dots & Y_{6} \\
    Y_{7} & Y_{8} & \dots  & Y_{12} \\
\end{bmatrix}.
\]
Considering each output pixel assigned to a neuron, Table \ref{FID} shows the convolutional process of this example in a way similar to the fully-connected layer computations, where input pixels are read sequentially at each clock cycle (CC) and the neurons share the same input pixels. This example considers $C_{in} = 1$, $C_{out} = 1$, $H_{in} = 2$, $W_{in} = 8$, $H_f = 1$, $W_f = 3$ and $S = 1$. Similar to the fully-connected dataflow, each neuron loads a different weight at each time step, subsequently accumulating the weighted input pixels. The number of time steps required to perform the convolutional computations is also equal to the number of input pixels, $H_{in} \times W_{in}$. When passed to the next neuron belonging to the same row of the output activation map, the weights need to be shifted of one position. However, weight passing between neurons of different rows requires a shift of $W_f$ positions, as can be observed between output $\#6$ and $\#7$ in Table \ref{FID}.\par

\begin{table}[!t]
\renewcommand{\arraystretch}{1.3}
\renewcommand{\thefootnote}{\alph{footnote}}
\centering

\caption{The FID for Convolutional Computations.} % title of Table

\scalebox{0.5}{
\Large
\begin{tabular}{c c | c c c c c c : c c c c c c} % centered columns (4 columns)
\hline %inserts double horizontal lines
&  & \multicolumn{6}{c:}{1$^{st}$ row of output activation map} & \multicolumn{6}{c}{2$^{nd}$ row of output activation map}\\
\hline
\multirow{2}{*}{CC} & \multirow{2}{*}{Inputs} &  \multicolumn{12}{c}{Outputs} \\
&  &  \#1  &  \#2 &  \#3 &  \#4 &  \#5 &  \#6  &  \#7  &  \#8 &  \#9 &  \#10 &  \#11 &  \#12 \\
\hline
 \#1 & $X_1\times$ & \colorbox{black!10}{$W_1$}  & $0$ & $ 0$ & $ 0$ & $ 0$ & $ 0$ & $ 0$ & $ 0$ & $ 0$ & $ 0$ & $ 0$ & $ 0$ \\

 \#2 & $X_2\times $ & \colorbox{black!10}{$ W_2$}  & \colorbox{black!30}{$W_1$} & $ 0$ & $ 0$ & $0$ & $0$  & $0$ & $ 0$ & $ 0$ & $ 0$ & $ 0$ & $ 0$\\

 \#3 & $X_3\times $ & \colorbox{black!10}{$W_3$}  & \colorbox{black!30}{$ W_2$} & \colorbox{black!50}{$ W_1$} & $ 0$ & $ 0$ & $ 0$  & $ 0$ & $ 0$ & $ 0$ & $ 0$ & $ 0$ & $ 0$\\

 \#4 & $X_4\times$ & $ 0$  & \colorbox{black!30}{$ W_3$} & \colorbox{black!50}{$ W_2$} & \colorbox{black!10}{$ W_1$} & $ 0$ & $ 0$  & $ 0$ & $ 0$ & $ 0$ & $ 0$ & $ 0$ & $ 0$\\

 \#5 & $X_5\times $ & $ 0$  & $ 0$ & \colorbox{black!50}{$ W_3$} & \colorbox{black!10}{$ W_2$} & \colorbox{black!30}{$ W_1$} & $ 0$  & $ 0$ & $ 0$ & $ 0$ & $ 0$ & $ 0$ & $ 0$\\

 \#6 & $X_6\times $  & $ 0$  & $0$ & $0$ & \colorbox{black!10}{$W_3$} & \colorbox{black!30}{$W_2$} & \colorbox{black!50}{$W_1$}  & $0$ & $0$ & $0$ & $0$ & $0$ & $0$\\

 \#7 & $X_7\times$  & $0$  & $0$ & $0$ & $0$ & \colorbox{black!30}{$ W_3$} & \colorbox{black!50}{$W_2$}  & $0$ & $0$ & $0$ & $0$ & $0$ & $0$\\

 \#8 & $X_8\times$ & $0$  & $0$ & $0$ & $0$ & $ 0$ & \colorbox{black!50}{$ W_3$}  & $ 0$ & $ 0$ & $ 0$ & $ 0$ & $ 0$ & $ 0$\\

%\hdashline

 \#9 & $X_9\times$ & $0$ & $ 0$ & $0$ & $0$ & $0$ & $ 0$ & \colorbox{black!10}{$ W_1$}  & $ 0$ & $ 0$ & $ 0$ & $ 0$ & $0$ \\

 \#10 & $X_{10}\times$ &  $0$ & $ 0$ & $ 0$ & $ 0$ & $ 0$ & $ 0$ & \colorbox{black!10}{$ W_2$}  & \colorbox{black!30}{$ W_1$} & $0$ & $ 0$ & $ 0$ & $ 0$ \\

 \#11 & $X_{11}\times$ &  $ 0$ & $ 0$ & $ 0$ & $ 0$ & $ 0$ & $0$ & \colorbox{black!10}{$ W_3$}  & \colorbox{black!30}{$W_2$} & \colorbox{black!50}{$ W_1$} & $ 0$ & $0$ & $0$\\

 \#12 & $X_{12}\times $ &  $ 0$ & $0$ & $0$ & $0$ & $ 0$ & $0$ & $ 0$  & \colorbox{black!30}{$W_3$} & \colorbox{black!50}{$W_2$} & \colorbox{black!10}{$W_1$} & $0$ & $0$\\

 \#13 & $X_{13}\times $ & $0$ & $0$ & $0$ & $0$ & $0$ & $0$ & $0$  & $0$ & \colorbox{black!50}{$W_3$} & \colorbox{black!10}{$W_2$} & \colorbox{black!30}{$W_1$} & $0$ \\

 \#14 & $X_{14}\times $ &  $0$ & $0$ & $0$ & $0$ & $ 0$ & $ 0$ & $0$  & $0$ & $0$ & \colorbox{black!10}{$ W_3$} & \colorbox{black!30}{$W_2$} & \colorbox{black!50}{$ W_1$} \\

 \#15 & $X_{15}\times $ &  $0$ & $ 0$ & $ 0$ & $ 0$ & $ 0$ & $ 0$ & $ 0$  & $0$ & $0$ & $ 0$ & \colorbox{black!30}{$W_3$} & \colorbox{black!50}{$W_2$}\\

 \#16 & $X_{16}\times$  & $ 0$ & $ 0$ & $ 0$ & $ 0$ & $ 0$ & $ 0$ & $ 0$  & $ 0$ & $ 0$ & $ 0$ & $ 0$ & \colorbox{black!50}{$W_3$}\\
\hline
\multicolumn{2}{c}{$\sum$}  & $Y_{1}$ & $Y_{2}$ & $Y_{3}$ & $Y_{4}$ & $Y_{5}$ & $Y_{6}$ & $Y_{7}$  & $Y_{8}$ & $Y_{9}$ & $Y_{10}$ & $Y_{11}$ & $Y_{12}$\\
\end{tabular}}
\label{FID} % is used to refer this table in the text
\end{table}

A direct implementation of the convolutional process in Table \ref{FID} requires a large number of PEs, or neurons, each of them with a low utilization factor (UF). In \cite{TCAS} it was shown that 3 PEs, which are denoted by different colors in Table \ref{FID}, are sufficient to perform the convolutions. In fact, there are only 3 active neurons at each time step. Each PE thus receives its input at clock cycle $3 \times i + 1$, $3 \times i + 2$ and $3 \times i + 3$. Their outputs are also valid after 3 clock cycles in the given example. So far, we only considered a case with $H_f = 1$. In case of $H_f = 3$, the procedure in Table \ref{FID} has to be repeated 2 times more: the first iteration with $W_1$, $W_2$ and $W_3$, the second with $W_4$, $W_5$ and $W_6$, and the final one with $W_7$, $W_8$ and $W_9$. Similarly, for higher values of $C_{in}$, the process has to be to repeated $C_{in}$ times. Therefore, a memory is required to store the partial values generated by the 3 neurons for each output pixel. In general, $N$ output pixels can be computed using 3 neurons (i.e. PEs) and 3 separate $N/3$-element SRAM memories working in parallel. The unit generating the $N$ output pixels of an output activation map is referred to as a 1D tile. Parallel 1D tiles can be also exploited to generate $p$ out of $C_{out}$ output activation maps in parallel. Using $p$ parallel 1D tiles reduces both the latency and memory access by a factor of $p$. The input pixels are shared among all the $p$ 1D tiles.

\section{Generalized Fully-Connected Inspired Dataflow (GFID)}\label{sec:GFID}
Let us define a generalized form of the FID as a matrix $M$:

\begin{equation}
\small
M =
\begin{matrix}
\begin{bmatrix}
 \left.\begin{array}{ccccc}
 W_1    & 0        & 0       & \cdots & 0 \\[0.3em]
 W_2    & \rvdots  &         &        & \\[0.3em]
        & 0        & \rvdots &        & \\[0.3em]
\hdashline \\[-0.3em]
\rvdots & W_1      &         &        & \\[0.3em]
        & W_2      & 0       &        & \rvdots\\[0.3em]
 W_{W_f}&          & 0       &        & \\[0.3em]
 0      & \rvdots  & W_1     &\rvdots & \\[0.3em]
        &          & W_2     &        & \\[0.3em]
        & W_{W_f}  &         &        & 0\\[0.3em]
        & 0        & \rvdots &        & W_1\\[0.3em]
 \rvdots&          &         &        & W_2\\[0.3em]
        & \rvdots  & W_{W_f} &        & \\[-0.2em]
        &          & \rvdots & \ddots & \rvdots\\[0.3em]
 0      & \cdots   & 0       & \cdots & W_{W_f}\\[0.3em]
 \end{array} \right.
\end{bmatrix},

\begin{matrix}
 \left.\begin{array}{l}
\MyLBrace{5ex}{S} \\[0.1em]
 \\[-0.3em]
\\[0.3em]
\\[0.3em]
\\[0.3em]
\\[0.3em]
\\[0.3em]
\\[0.3em]
\\[0.3em]
\\[0.3em]
\\[0.3em]
\\[0.3em]
\\[0.3em]
 \end{array}\right.
\end{matrix}
 \end{matrix}
\end{equation}
where each column of the matrix $M$ can contain only $W_f$ non-zero elements at most. The shift amount within each row of the output activation map is equal to $S$, denoted with a dashed line in the matrix $M$. The number of columns of the matrix $M$ indicates the $N$ output pixels that belong to the same row of the output activation map, while the number of rows of $M$ denotes the required number of clock cycles.\par

In this Section, we use the GFID matrix $M$ to represent different filter sizes used in the state-of-the-art CNNs (i.e., AlexNet, VGGNet and ResNet). AlexNet uses filter sizes of $11 \times 11$ with $S = 4$, $5 \times 5$ with $S = 1$, and $3 \times 3$ with $S = 1$. The filter sizes used in VGGNets are fixed to $3 \times 3$ with $S = 1$. Finally, ResNets use filter sizes of $7 \times 7$ with $S = 2$, $3 \times 3$ with $S = 1$, and $1 \times 1$ with $S = 1$.

\subsection{Filters with $W_f = 3$ and $S = 1$}\label{GFID:3x3}
In Section \ref{pre:FID}, we showed that 3 PEs are sufficient to perform the convolutions for filter size of $3 \times 3$ with $S = 1$. Therefore, a 1D tile containing only 3 neurons can perform the convolutional computations. Considering a convolution of a row of a filter map with its corresponding input pixels, $N + 2$ clock cycles are required to generate $N$ output pixels which belong to the same row of the output activation map. For instance, in the given example in Table \ref{FID}, 8 clock cycles are required to generate the output pixels of the first row of the output activation map (i.e., the first 6 output pixels). This example can also be expressed using the GFID matrix $M$ as follows:
\begin{equation}
\small
M_{8\times 6} =
\begin{bmatrix}
 \left.\begin{array}{cccccc}
 W_1    & 0        & 0       & 0      & 0   & 0 \\
 W_2    & W_1      & 0       & 0      & 0   & 0  \\
 $\colorbox{black!50}{$W_3$}$    & $\colorbox{black!50}{$W_2$}$      & $\colorbox{black!50}{$W_1$}$     & 0      & 0   & 0  \\
 0      & W_3      & W_2     & W_1    & 0   & 0  \\
 0      & 0        & W_3     & W_2    & W_1 & 0  \\
 0      & 0        & 0       & W_3    & W_2 & W_1  \\
 0      & 0        & 0       & 0      & W_3 & W_2  \\
 0      & 0        &  0      &0       &  0  & W_3 \\
 \end{array} \right.
\end{bmatrix}.
\end{equation}
The matrix $M$ also confirms that there are only 3 active neurons at each time steps, highlighted in dark gray.

\subsection{Filters with $W_f = 5$ and $S = 1$}\label{GFID:5x5}
The convolutional computations for filters with $W_f = 5$ and $S = 1$ are performed in a way similar to the convolutional computations of the filters with $W_f = 3$ and $S = 1$, with the difference that $5$ neurons are active at each time step. Thus, a 1D tile with 5 PEs can perform the computations for this filter size. Moreover, $N + 4$ clock cycles are required to generate $N$ output pixels which belong to the same row of the output activation map.

%The GFID matrix of such filters, generating $N = 6$ output pixels, is obtained as
% \begin{equation}
% \small
% M_{10\times 6} =
% \begin{bmatrix}
%  \left.\begin{array}{cccccc}
%  W_1    & 0        & 0       & 0      & 0   & 0 \\
%  W_2    & W_1      & 0       & 0      & 0   & 0  \\
%  W_3    & W_2      & W_3     & 0      & 0   & 0  \\
%  W_4    & W_3      & W_2     & W_1    & 0   & 0  \\
%  $\colorbox{black!50}{$W_5$}$    & $\colorbox{black!50}{$W_4$}$      & $\colorbox{black!50}{$W_3$}$     & $\colorbox{black!50}{$W_2$}$    & $\colorbox{black!50}{$W_1$}$ & 0  \\
%  0      & W_5      & W_4     & W_3    & W_2 & W_1  \\
%  0      & 0        & W_5     & W_4    & W_3 & W_2  \\
%  0      & 0        &  0      & W_5    & W_4 & W_3 \\
%  0      & 0        &  0      & 0      & W_5 & W_4 \\
%  0      & 0        &  0      & 0      &  0  & W_5 \\
%  \end{array} \right.
% \end{bmatrix}.
% \end{equation}
%The matrix $M$ shows that there are only 5 active neurons at each time step, suggesting that a 1D tile with 5 PEs can perform the computations for this filter size. Moreover, $N + 4$ clock cycles are required to generate $N$ output pixels which belong to the same row of the output activation map.

\subsection{Filters with $W_f = 1$ and $S = 1$}\label{GFID:1x1}
The following matrix $M$ shows the convolutional computations for filters with $W_f = 1$ and $S = 1$.

\begin{equation}
\small
M_{5\times 5} =
\begin{bmatrix}
 \left.\begin{array}{cccccc}
 $\colorbox{black!50}{$W_1$}$    & 0        & 0       & 0      & 0   \\
 0      & W_1      & 0       & 0      & 0    \\
 0      & 0        & W_1     & 0      & 0    \\
 0      & 0        & 0       & W_1    & 0   \\
0       & 0        & 0       & 0      & W_1 \\
 \end{array} \right.
\end{bmatrix}.
\end{equation}
Contrary to other filter sizes, its GFID matrix $M$ is square: the number of clock cycles required to generate $N$ output pixels is equal to $N$. As denoted in the matrix $M$, there is only one active neuron at each clock cycle. Consequently, its 1D tile requires only one PE to perform the convolutional computations.

\subsection{Filters with $W_f = 7$ and $S = 2$}\label{GFID:7x7}
So far, we only considered a stride value $S=1$. However, both AlexNet and ResNet contain layers computing convolutions with $S \ge 1$. Considering filters with $W_f = 7$ and $S = 2$, the shift amounts within each row of the output activation map is equal to 2 as shown in the following matrix $M$:
\begin{equation}
\small
M_{15\times 5} =
\begin{bmatrix}
 \left.\begin{array}{cccccc}
 W_1    & 0        & 0       & 0      & 0   \\
 W_2    & 0        & 0       & 0      & 0    \\
 W_3    & W_1      & 0       & 0      & 0    \\
 W_4    & W_2      & 0       & 0      & 0   \\
 W_5    & W_3      & W_1     & 0      & 0   \\
 W_6    & W_4      & W_2     & 0      & 0   \\
 $\colorbox{black!50}{$W_7$}$     & $\colorbox{black!50}{$W_5$}$       & $\colorbox{black!50}{$W_3$}$      & $\colorbox{black!50}{$W_1$}$     & 0   \\
 0      & W_6      & W_4     & W_2    & 0   \\
 0      & W_7      & W_5     & W_3    & W_1 \\
 0      & 0        & W_6     & W_4    & W_2 \\
 0      & 0        & W_7     & W_5    & W_3 \\
 0      & 0        & 0       & W_6    & W_4 \\
 0      & 0        & 0       & W_7    & W_5 \\
 0      & 0        & 0       & 0      & W_6 \\
 0      & 0        & 0       & 0      & W_7 \\
 \end{array} \right.
\end{bmatrix}.
\end{equation}
While the higher stride value linearly decreases the number of pixels in the output activation maps, it also reduces the number neurons required to perform the convolutional computations. For instance, the above matrix $M$ shows that there are only $4$ active neurons at each time step, while the width of the filter $W_f=7$. According to the matrix $M$, $15$ clock cycles are required to generate $5$ output pixels in the given example.

\subsection{Filters sizes with $W_f = 11$ and $S = 4$}\label{GFID:11x11}
The matrix $M$ for filters with $W_f = 11$ and $S = 4$ is as follows:
\begin{equation}
\small
M_{23\times 4} =
\begin{bmatrix}
 \left.\begin{array}{ccccc}
 W_1    & 0        & 0       & 0       \\
 W_2    & 0        & 0       & 0       \\
 W_3    & 0        & 0       & 0       \\
 W_4    & 0        & 0       & 0        \\
 W_5    & W_1      & 0       & 0       \\
 W_6    & W_2      & 0       & 0       \\
 W_7    & W_3      & 0       & 0       \\
 W_8    & W_4      & 0       & 0       \\
 W_9    & W_5      & W_1     & 0        \\
 W_{10} & W_6      & W_2     & 0       \\
 $\colorbox{black!50}{$W_{11}$}$ & $\colorbox{black!50}{$W_7$}$      & $\colorbox{black!50}{$W_3$}$     & 0       \\
 0      & W_8      & W_4     & 0        \\
 0      & W_9      & W_5     & W_1      \\
 0      & W_{10}   & W_6     & W_2      \\
 0      & W_{11}   & W_7     & W_3      \\
 0      & 0        & W_8     & W_4      \\
 0      & 0        & W_9     & W_5     \\
 0      & 0        & W_{10}  & W_6    \\
 0      & 0        & W_{11}  & W_7     \\
 0      & 0        & 0       & W_8     \\
 0      & 0        & 0       & W_9     \\
 0      & 0        & 0       & W_{10}  \\
 0      & 0        & 0       & W_{11}  \\
 \end{array} \right.
\end{bmatrix}.
\end{equation}
Despite of the large width of the filter, the number of active neurons at each time step is only 3, thanks to the large stride value. However, the number of clock cycles required to generate 4 output pixels is $23$, which is rather high and can result in a long latency.

\subsection{Utilization Factor for Different Filter Sizes}\label{GFID:UF}
As discussed in Section \ref{pre:FID}, the number of clock cycles required to perform the convolutions using FID is equal to the number of input pixels, and it is the same for GFID. Considering $C_{in} = 1$ and $H_f = 1$, in order to generate $N$ pixels of an output activation map, $S \times N + W_f -S$ clock cycles are required to perform the convolutions according to Eq. (\ref{eq_conv}). Let us define the number of required PEs in the 1D tile as $T$. The number of pixels computed by each neuron is equal to $N/T$ when $N$ is a multiple of $T$. Each neuron also requires $W_f$ clock cycles to generate an output pixel. Therefore, the utilization factor of GFID can be expressed as
\begin{equation}
\text{UF} = \dfrac{\dfrac{N}{T} \times W_f}{S \times N + W_f -S} \times 100.
\label{UF}
\end{equation}\par
In Section \ref{intro}, we discussed the importance of high performance efficiency. The utilization factor of PEs in a convolutional accelerator is also linearly proportional to its performance efficiency. Any increasing in the utilization factor of PEs exploited in the 1D tile results in an increase in performance efficiency. Considering the fact that $W_f$ and $S$ are usually small, a high UF is achieved for a large value of $N$. In other word, the maximum achievable utilization factor can be obtained as
\begin{equation}
\text{UF$_{max}$} = \lim_{N\to\infty} \text{UF}  = \dfrac{W_f}{T \times S} \times 100. \label{UFmax}
\end{equation}

Eq. (\ref{UFmax}) suggests that the highest performance efficiency is obtained when $N \gg (W_f - S)$. The maximum utilization factors for filters with [$W_f, S$] equal to [1, 1], [3, 1], [5, 1], [7, 2] and [11, 4] are 100\%, 100\%, 100\%, 88\% and 92\%, respectively, showing the high performance efficiency of the proposed GFID.

\section{Multi-Mode Inference Engine}\label{sec:MMIE}

\begin{table}[!t]
\renewcommand{\arraystretch}{1.3}
\renewcommand{\thefootnote}{\alph{footnote}}
\caption{Breakdown of Number of PEs Required Per Tile}
\centering
\small
\scalebox{1}{
\begin{tabular}{c c c c c}
\hline\hline
\multicolumn{1}{c}{Network} & \multicolumn{1}{c}{$H_f \times W_f$} & $S$ & $T$ & \# layers\\
\hline
\hline
\multirow{3}{*}{AlexNet \cite{alexnet}}  & $11\times11$ & 4& 3 & 1 out of 5 \\
  & $5\times5$ & 1 & 5 & 1 out of 5 \\
    & $3\times3$ & 1 & 3 & 3 out of 5 \\
\hline
\multirow{2}{*}{ResNet-50 \cite{resnet}}  & $7\times7$ & 2 & 4 & 1 out of 49 \\
  & $3\times3$ & 1 & 3 & 16 out of 49 \\
    & $1\times1$ & 1 & 1 & 32 out of 49 \\
\hline
\multirow{1}{*}{VGG-16 \cite{vgg}}  & $3\times3$ & 1 & 3 & 13 out of 13 \\
\hline
\hline
\end{tabular}}
\label{PE_number}
\end{table}

In Section \ref{sec:GFID}, we showed that different filter sizes require different number of PEs per tile. Table \ref{PE_number} summarizes the number of required PEs per tile for each layer of AlexNet, VGGNet-16 and ResNet-50. AlexNet consists of 5 layer of convolutions with filter sizes of $11 \times 11$, $5 \times 5$ and $3 \times 3$. Performing the GFID on the AlexNet layers show that 4 out of 5 layers (i.e., the layers with filter sizes of $11 \times 11$ and $3 \times 3$) only require 3 PEs to perform the computations, while the remaining layer requires 5 PEs per tile. Therefore, $T = 3$ is the most frequent number in AlexNet for convolutional processes. The filter size and stride are fixed to $3 \times 3$ and one pixel for the convolutional computations of VGGNets \cite{vgg}, respectively: as a result, the whole computations of VGGNets can be performed using 3 PEs per tile. There are different VGGNet models in literature: in this paper, we use VGGNet-16, which contains 13 convolutions and 3 fully-connected layers, for experimental purposes. Similar to VGGNets, ResNets also come in different flavors. The first layer of ResNets is fixed to the receptive field of $7 \times 7$ and stride of $S = 2$. The filter sizes of the remaining layers are either fixed to $3 \times 3$ (for ResNet-18 and ResNet-34) or a combination of $1 \times 1$ and $3 \times 3$ (for ResNet-50, ResNet-101 and ResNet-152) \cite{resnet}. Therefore, the dominant filter sizes are $1 \times 1$ and $3 \times 3$, which require one and 3 PEs per tile to perform the convolutional computations, respectively. In Table \ref{PE_number}, we report the requirements for ResNet-50.

\begin{figure}[!t]
\centering
\includegraphics[scale = 0.37]{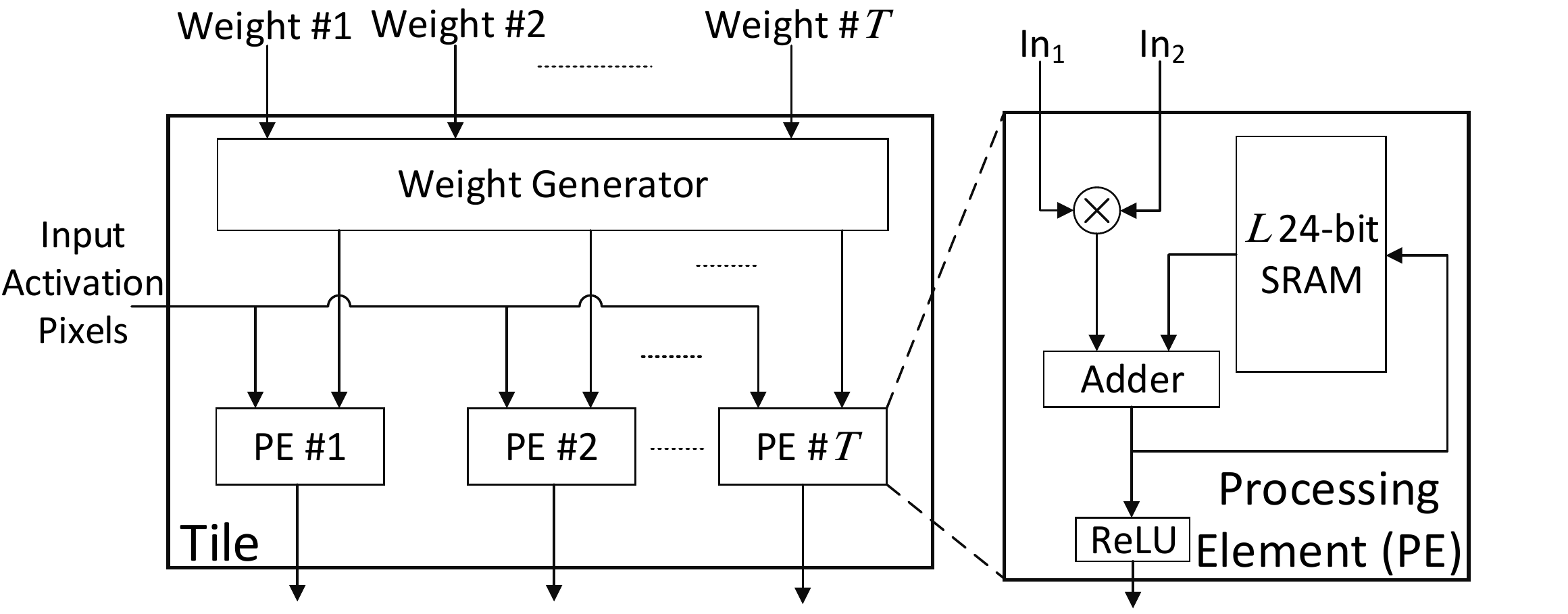}
\caption{The high-Level Architecture of a 1D Reconfigurable Tile.}
\label{Tile}
\end{figure}

Fig. \ref{Tile} shows the high level architecture of the 1D tile. It consists of two main sub-blocks: the weight generator and $K$ PEs working in parallel. All the PEs share the same input activation pixel while their weights are different. Each PE takes an input activation map and its corresponding weight according to the proposed GFID and performs the accumulation-multiplication for the first row of the first input filter, i.e., $W_1$, $W_2$, $\dots$, $W_{W_f}$. This process takes $W_f$ clock cycles and the computed partial value is stored in a memory of $L$ elements. Afterwards, the PE starts the processing of another output activation pixel, using the same weights. The convolutional computations of the first row of the first input filter require $S \times N + W_f - S$ clock cycles, as discussed in Section \ref{GFID:UF}. Upon reaching this point, the partial value of the first output activation pixel is read from the memory and the computations of the second row of the first input filter are performed for $S \times N + W_f -S$ clock cycles. In general, this procedure is repeated for $H_f$ times until the computations of the first filter are finished (i.e., upon completion of $H_f \times (S \times N + W_f -S)$ clock cycles). At this point, the computation of the second of the $C_{in}$ filters starts. Upon completion of $C_{in} \times H_f \times (S \times N + W_f -S)$ clock cycles, the output value of each PE is passed through the ReLU and the result is stored in the off-chip memory.

So far, we introduced a high-level architecture for the 1D tile and explained the high level procedure of convolutional computations. In order to perform the computations while achieving a high performance efficiency, the number of PEs per tile has to be reconfigurable. In order words, $K$ instantiated PEs have to dynamically adapt to act as a multiple of $T$ PEs to achieve the maximum possible utilization factor. The closed form solution for this strategy is
\begin{equation}
K = \text{LCM}(T_i),~~i \in \{1,3,4,5\},
\end{equation}
where LCM denotes the least common multiple. Using this approach, 60 PEs are required to achieve the maximum possible utilization factor for all the network sizes listed in Table \ref{PE_number}. Depending on the required $T$, the 60 PEs can dynamically behave as a set of $T$ PEs. For instance, they can act as 60, 20, 15 and 12 parallel tiles for $T$ equal to 1, 3, 4 and 5, respectively, where each tile also contains 1, 3, 4 and 5 PEs. However, using 60 reconfigurable PEs is not trivial and results in a complex address generator.\par

Table \ref{PE_number} shows that $T = 1$ and $T = 3$ are the dominant minimum numbers of PEs for the three well-known CNNs. More precisely, the two filters with $W_f = 5$ and $W_f = 7$ have the least impact on the overall performance efficiency of CNNs, since they are used in only one layer of CNNs. Therefore, we use $K=6$ PEs inside the reconfigurable tile: the reason is twofold. First of all, 6 PEs can be easily used as 2 and 6 tiles containing 3 and 1 PEs for $T = 3$ and $T = 1$, which are the dominant minimum numbers of PEs for the three well-known CNNs. Secondly, they can perform the computations for $T = 4$ and $T = 5$ with a minimum level of complexity for the address generator unit. In this case, with $K$ larger than what strictly necessary, the number of clock cycles required to perform the convolutional computations remains the same. However, the utilization factors of PEs for these cases decreases.

\subsection{Reconfigurable Weight Generator Unit}\label{MMIE:WG}

The weight generator unit provides each neuron an appropriate weight according to the proposed GFID. The weight generator unit consists of 6 register sets where each set contains 11 registers. The appropriate weight for each neuron is provided by selecting among these shift registers.

\begin{figure*}[!t]
    \centering
    \subfigure[]{
        \includegraphics[scale = 0.35]{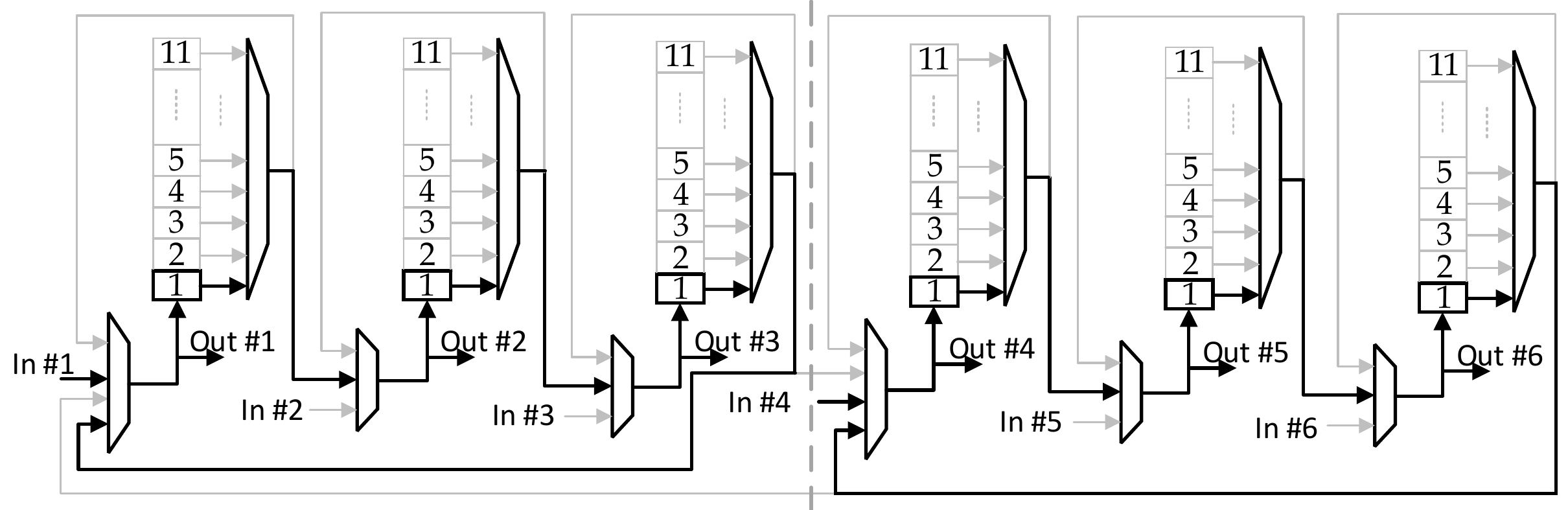}
        \label{WG_3x3}
    }
    \subfigure[]{
        \includegraphics[scale = 0.35]{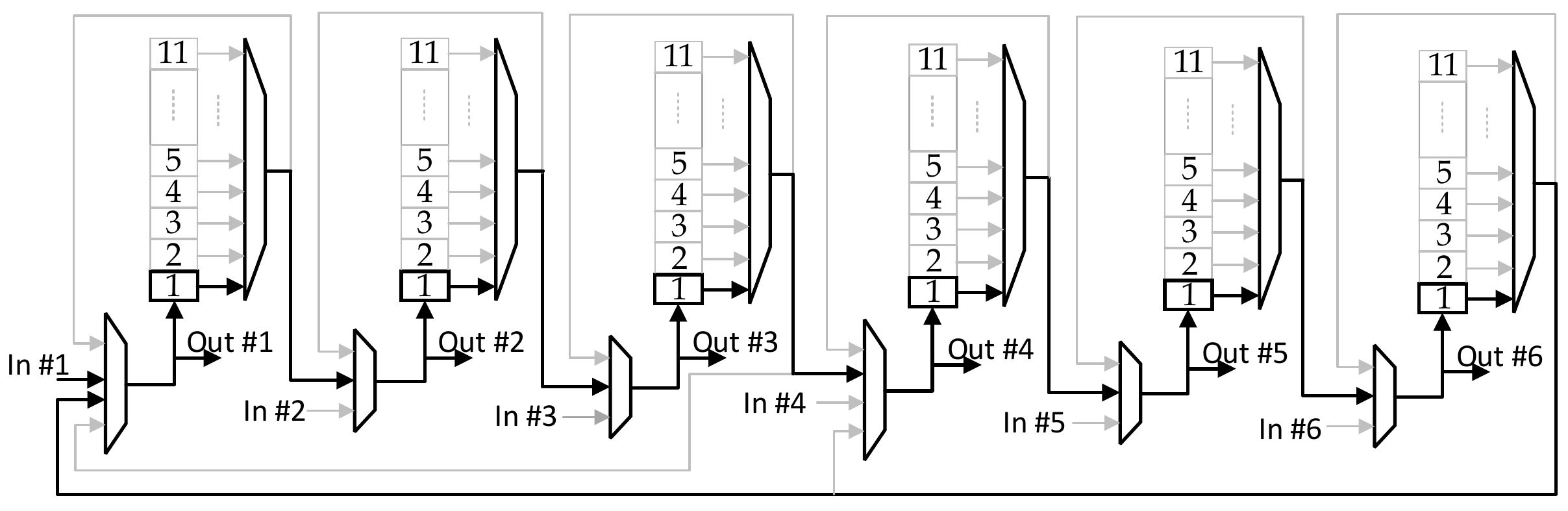}
        \label{WG_5x5}
    }
    \subfigure[]{
        \includegraphics[scale = 0.35]{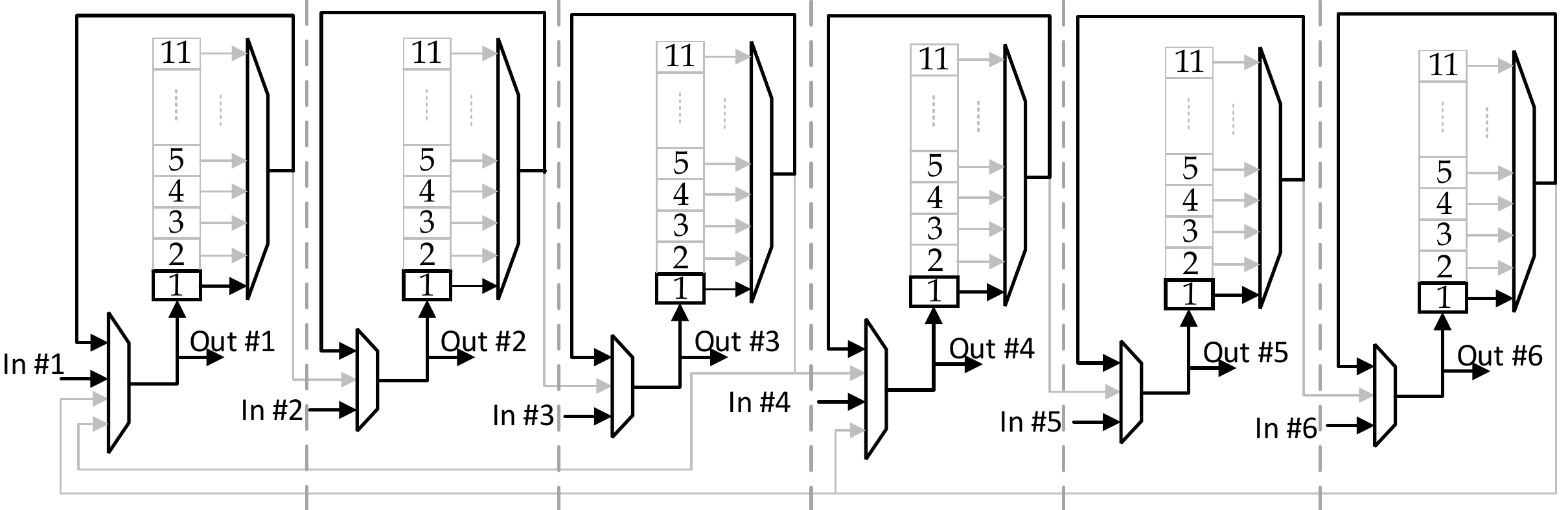}
        \label{WG_1x1}
    }
    \subfigure[]{
        \includegraphics[scale = 0.35]{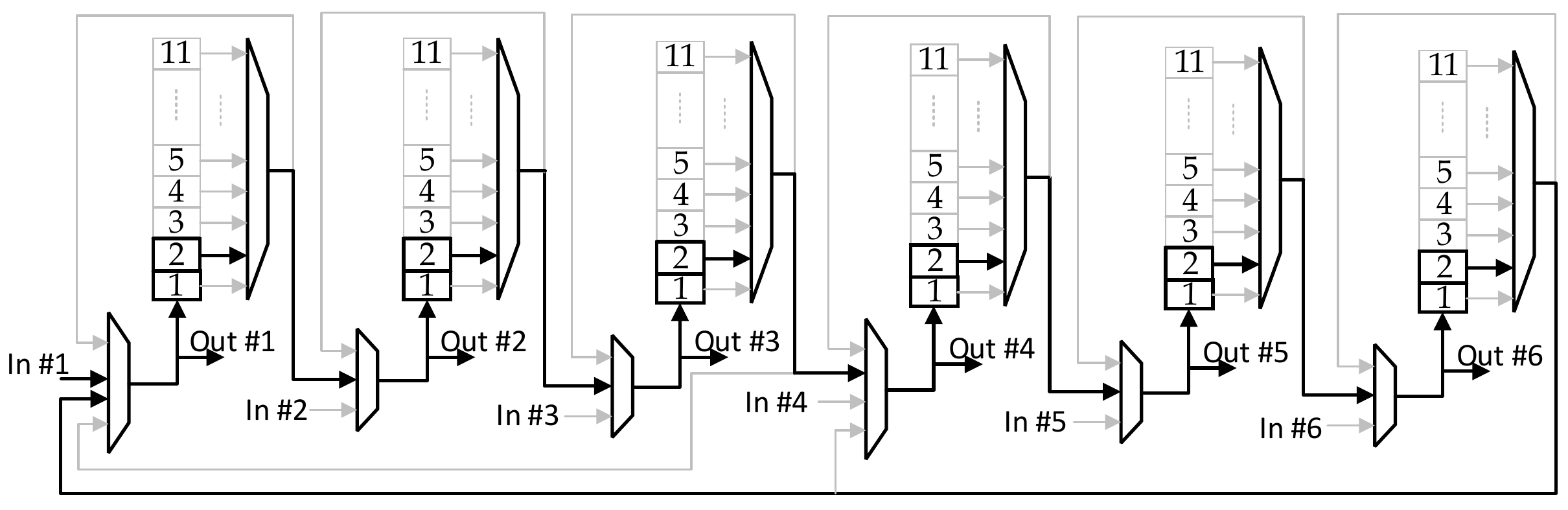}
        \label{WG_7x7}
    }
    \subfigure[]{
        \includegraphics[scale = 0.35]{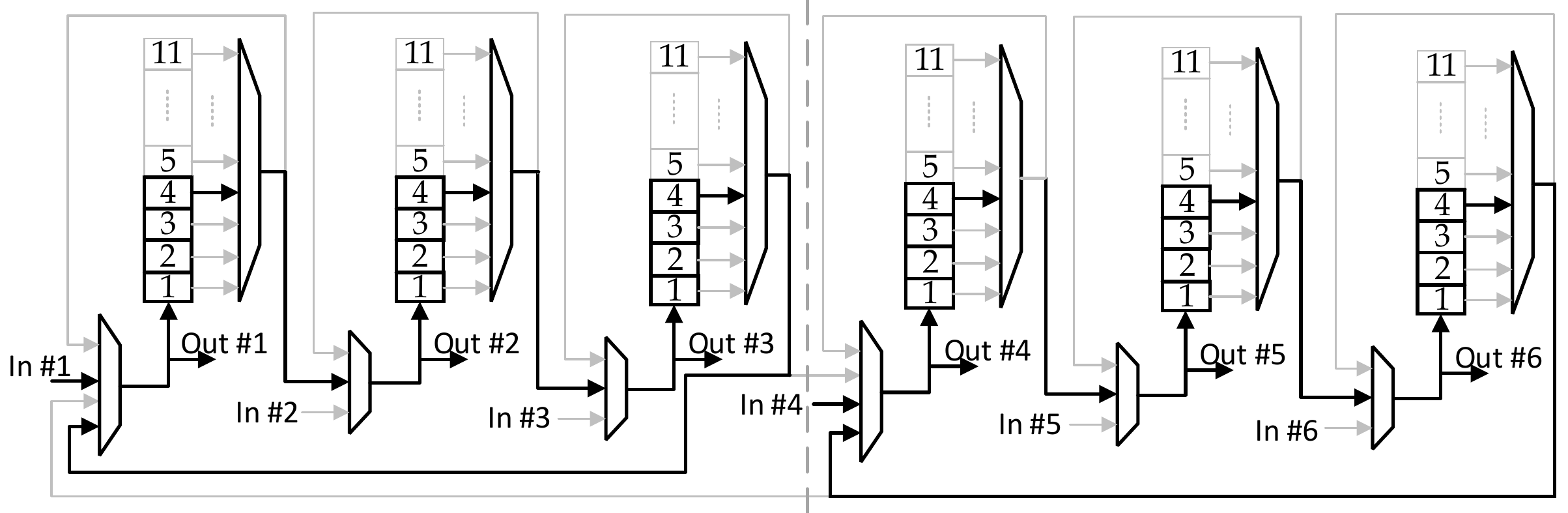}
        \label{WG_11x11}
    }
    \subfigure[]{
        \includegraphics[scale = 0.35]{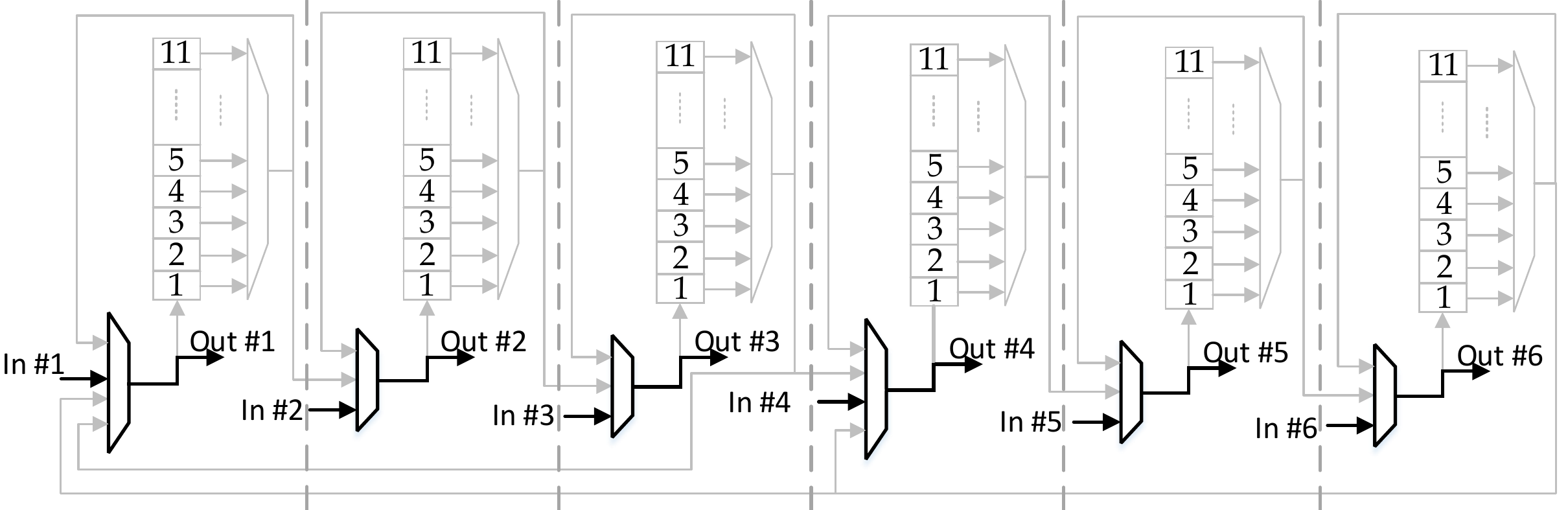}
        \label{WG_fc}
    }
    \caption{Involved hardware resources and paths in case of convolution computations for (a) $W_f = 3$ and $S = 1$, (b) $W_f = 5$ and $S = 1$, (c) $W_f = 1$ and $S = 1$, (d) $W_f = 7$ and $S = 2$, (e) $W_f = 11$ and $S = 4$, and (f) fully-connected computations.}
    \label{mult}
\end{figure*}

\subsubsection{Filters With $W_f = 3$ and $S = 1$}\label{MMIE:WG3x3}
As discussed in Section \ref{sec:MMIE}, in case of $W_f = 3$ and $S = 1$, the 1D reconfigurable tile containing 6 neurons can function as two tiles of 3 neurons each. Fig. \ref{WG_3x3} shows the weight generator unit and its working path highlighted in black when using $W_f = 3$ and $S = 1$. It is worth noting that tiles are separated using a dashed line. Each tile loads the weights of the first row of the first filter (i.e., $W_1$, $W_2$ and $W_3$) through the input ports denoted as In \#1 and In \#2 in Fig. \ref{WG_3x3}. These weights then loop through the first register of each set to provide one clock cycle delay for each neuron according to (\ref{GFID:3x3}). Considering Eq. (\ref{UF}), the utilization factor of each neuron for this case can be computed as
\begin{equation}
\text{UF} = \dfrac{N}{N + 2},
\end{equation}
which approaches 100\% for large values of $N$.

\subsubsection{Filters With $W_f = 5$ and $S = 1$}\label{MMIE:WG5x5}

In case of $W_f = 5$ and $S = 1$, we use 6 neurons to perform the convolutional processes while we showed that the minimum required number of neurons is 5 for this case (see Section \ref{sec:MMIE}). Therefore, the reconfigurable tile works as a single tile containing 6 PEs as shown in Fig. \ref{WG_5x5}. The tile takes the first row of the first filter (i.e., $W_1$, $W_2$, $\dots$, and $W_5$) through the input port denoted as In \#1. It then provides the required one clock cycle delay for each PE by passing the weights through the first register of each register set as highlighted in black in Fig. \ref{WG_5x5}. It is worth noting that 6 registers are used in this paradigm while only 5 of them required to store the weights. Therefore, the value of one register among the 6 registers is always zero to cancel out its effect on the computations. More precisely, we can assume the 5 weights (i.e., $W_1$, $W_2$, $\dots$, and $W_5$) as a set of 6 weights in which one of them is zero (i.e., $W_1$, $W_2$, $\dots$, $W_5$ and $0$). The utilization factor of each PE is also can be expressed as
\begin{equation}
\text{UF} = \dfrac{5N}{6N + 24}.
\end{equation}
In fact, using 6 neurons to perform the convolutions of $W_f = 5$ reduces the maximum achievable utilization factor from 100\% to 83\%.

\subsubsection{Filters With $W_f = 1$ and $S = 1$}\label{MMIE:WG1x1}

In Section \ref{GFID:1x1}, we showed that only one PE is sufficient to perform the computations for $W_f = 1$ and $S = 1$. Therefore, the reconfigurable 1D tile can be used as 6 parallel tiles, as depicted in Fig. \ref{WG_1x1}. The 6 tiles are separated using dashed lines and the involved hardware units and paths are highlighted in black. Each tile takes its weight (i.e., $W_1$) at the first clock cycle through the input ports In \#1 to In \#6. Afterwards, the imported weight loops through each tile and the first register of each register set. The utilization factor of each PE is equal to 100\% regardless of $N$, according to (\ref{UF}).

\subsubsection{Filters With $W_f = 7$ and $S = 2$}\label{MMIE:WG7x7}
Similar to the case of $W_f = 5$ and $S = 1$, 6 PEs are used to compute the convolutions for $W_f = 7$ and $S = 2$, while only 4 neurons are sufficient. As a result, the reconfigurable tile functions as a single tile containing 6 PEs (see Fig. \ref{WG_7x7}). The tile loads the weights of the first row of the first filter (i.e., $W_1$, $W_2$, $\dots$, and $W_7$) through the input port In \#1 and they loop through the black paths in Fig. \ref{WG_7x7}. In this scheme, the first two registers of each register set are used to provide the required two delays for each PE, as shown in (\ref{GFID:7x7}). It is worth mentioning that while 12 registers are used in this case, only 7 of them contain the weights. The utilization factor for this configuration is computed as follows:
\begin{equation}
\text{UF} = \dfrac{7N}{12N + 30}.
\end{equation}
Since 4 PEs are sufficient to perform the computations of this case, using 6 neurons highly affects the utilization factor and results in 53\% for large values of $N$. However, the final impact of this case when considering the computations of whole system is negligible due to the fact that this configuration is only used for one layer out of 49 in ResNet-50.

\subsubsection{Filters With $W_f = 11$ and $S = 4$}\label{MMIE:WG11x11}
Similar to the case with $W_f = 3$ and $S = 1$, 3 PEs are sufficient to perform the convolutional processes when using $W_f = 11$ and $S = 4$. Therefore, the reconfigurable tile functions as two tiles where each contains 3 PEs. The weights of the first row of the first filter (i.e., $W_1$, $W_2$, $\dots$, and $W_{11}$) are passed through input ports In \#1 and In \#4 to each tile, as shown in Fig. \ref{WG_11x11}. Since a stride value of 4 is used, the first four registers of each register set are used to provide the required four clock cycle delays (\ref{GFID:11x11}). A total of 12 registers are used in each tile while only 11 weights exist. Therefore, the remaining register is zero. The utilization factor of this case is also computed as
\begin{equation}
\text{UF} = \dfrac{11N}{12N + 21},
\end{equation}
achieving up to 92\% for large values of $N$.

\subsubsection{Fully-Connected Computations}\label{MMIE:WG_fc}
As discussed in Section \ref{pre}, semi-parallel architectures are a common approach to implement fully-connected layers, where the computations of each neuron are performed serially. For instance, considering a single neuron with 512 inputs (i.e., $n = 512$ and $m = 1$), 512 clock cycles are required to perform the computations of (\ref{fc_comp}) using a single PE. We can perform the computations of multiple neurons by instantiating multiple PEs in parallel as discussed in \cite{TCAS}. In this way, each PE shares the same input pixels while loading different weights. This approach can be easily realized using the proposed reconfigurable tile as illustrated in Fig. \ref{WG_fc}. In fact, the reconfigurable tile passes the incoming 6 parallel weights directly to each PE through multiplexers highlighted in black. The utilization factor of PEs for fully-connected computations is 100\%.

\subsection{Handling Weight Passing}\label{MMIE:WP}
So far, we discussed both the convolutional and fully-connected computations while not considering the weight passing cases for the sake of simplicity. However, weight passing occurrence is inevitable in convolutional computations and impacts both the processing time and utilization factor of PEs. Weight passing occurs when a tile performs the computations of more than one row of the output activation map. In this case, the weight passing from a neuron of a row to a neuron of another row takes $W_f$ clock cycles regardless of the stride value $S$, resulting in a longer latency and consequently a lower utilization factor for PEs. The total number of weight passing occurrences for computations of a single convolutional layer is equal to $H_{out} - 1$. We considered 11 registers for each register set to support the weight passing delay up to 11 clock cycles. Therefore, in case of weight passing in any of PEs, its corresponding register set provides the required delay depending on $W_f$.

\subsection{Exploiting Parallel Tiles}\label{MMIE:PT}
While the proposed reconfigurable tile can efficiently performs both fully-connected and convolutional computations, using a single tile results in a long latency and numerous memory accesses, as discussed in \cite{TCAS}. To address this issue, $p$ tiles are instantiated in parallel to generate multiple activation maps in parallel. Since the reconfigurable tile itself can function as up to 6 parallel tiles, the upper bound for the maximum number of tiles is $6p$ in MMIE. Therefore, the computational latency of MMIE is effectively reduced by a $p$ factor when compared to a single reconfigurable tile. Moreover, the memory accesses are reduced as well, since the input pixels are shared among the parallel tiles (see Fig. \ref{CL}), while each tile is fed by a different set of weights.

Exploiting parallel tiles requires an input bandwidth of $(1 + 6 \times P) \times 16$ bits ($6 \times p \times 16$ for weights and 16 for input pixels). However, most of the embedded and mobile devices cannot provide such a high bandwidth. To overcome this problem, MMIE leverages the pipelining technique first introduced in \cite{TCAS}. As discussed in Section \ref{sec:GFID}, each input pixel is read at each clock cycle while $W_f$ weights are read only for the first $W_f$ clock cycles when performing the convolutional process of the first row of the first input filter. The parameter $W_f$ is also a small value compared to the processing time of convolutions for the first row of the first input filter (i.e., $W_f \ll (S \times N + W_f - S)$). More precisely, the input bandwidth from $(W_f)^{th}$ clock cycle to $(S \times N + W_f - S)^{th}$ clock cycle is only occupied with input pixels. Therefore, we can fill out this available bandwidth by pipelining the tiles with up to $\lfloor (S \times N + W_f - S)/W_f \rfloor$ stages, while the additional latency overhead is negligible compared to the overall latency of the system.

\subsection{Processing Time and Memory Accesses of MMIE}\label{MMIE:PTMA}
\subsubsection{Convolutional Processes}\label{MMIE:PTMA:cp}
Earlier in Section \ref{sec:MMIE} we showed that in convolutional processes, a single tile computes $N$ out of $H_{out} \times W_{out}$ pixels of one of $C_{out}$ output activation maps within $C_{in} \times H_f \times (S \times N + W_f - S)$ clock cycles. We also showed that the total number of weight passing occurrences for the computation of a single convolutional layer is equal to $H_{out} - 1$, which causes additional $(W_f - 1) \times (H_{out} - 1)$ clock cycles for the computations of each row of the input filters. Considering $p$ parallel tiles, the number of required clock cycles is expressed as
\begin{align}
\text{CC} & =  \dfrac{W_{out} \times H_{out}}{N} \times (S \times N + W_f - S) \times H_f \times C_{in} \times \left \lceil \dfrac{C_{out}}{p} \right \rceil  \nonumber \\
 			  & + (W_f - 1) \times (H_{out} - 1) \times H_f \times C_{in} \times \left \lceil \dfrac{C_{out}}{p} \right \rceil.
 			  \label{cc}
\end{align}

\begin{table}[!t]
\renewcommand{\arraystretch}{1.3}
\renewcommand{\thefootnote}{\alph{footnote}}
\caption{The effective value of $N$ and $p$ in MMIE depending on $W_f$ and $S$}
\centering
\small
\scalebox{1}{
\begin{tabular}{c c c c c}
\hline\hline
\multicolumn{1}{c}{$H_f \times W_f$} & $S$  & $N_{eff}$ & $p_{eff}$\\
\hline
\hline
  $11\times11$ & 4 & 192 & 64 \\
  $7\times7$ & 2 & 384 & 32 \\
  $5\times5$ & 1 & 384 & 32 \\
  $3\times3$ & 1 & 192 & 64  \\
  $1\times1$ & 1 & 64 & 192 \\
\hline
\hline
\end{tabular}}
\label{eff_par}
\end{table}

\begin{figure*}[!t]
\centering
\subfigure[]{
        \includegraphics[scale = 0.6]{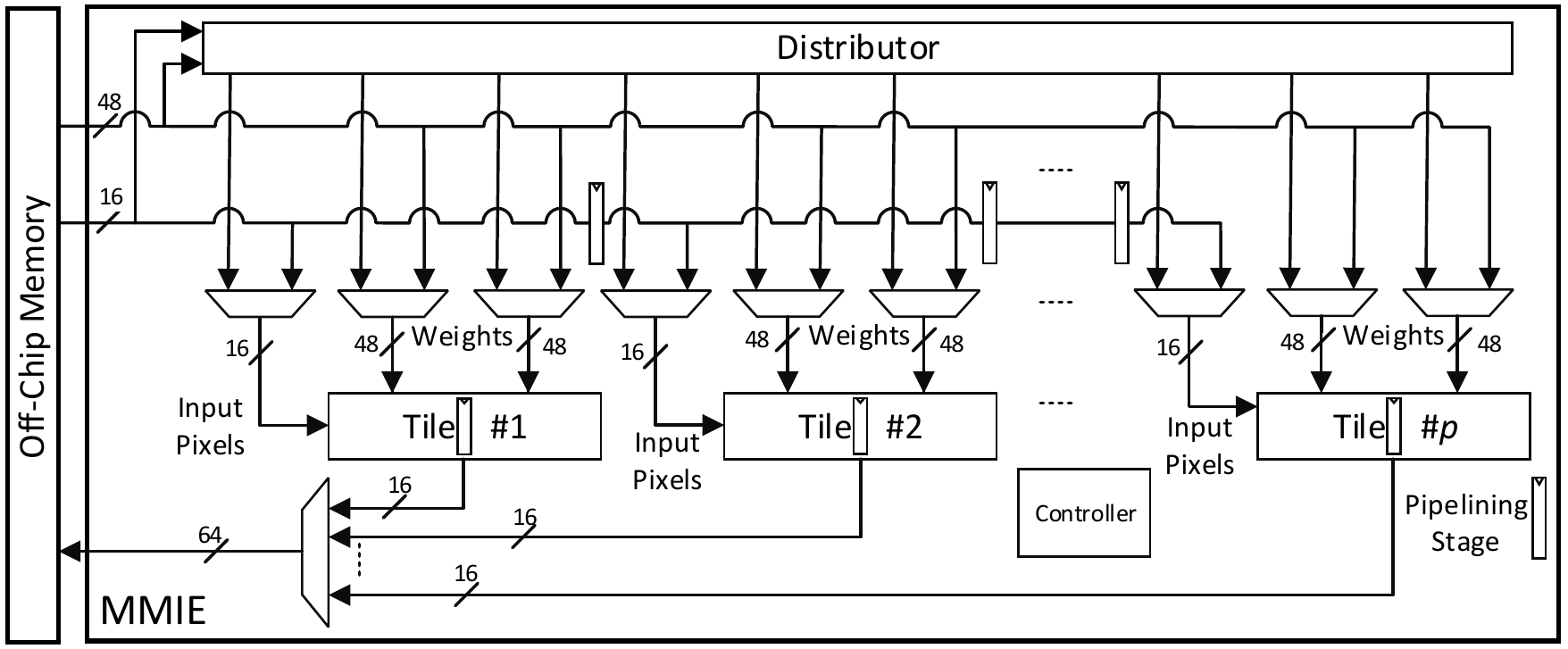}
        \label{MMIE}
    }
    \subfigure[]{
        \includegraphics[scale = 0.25]{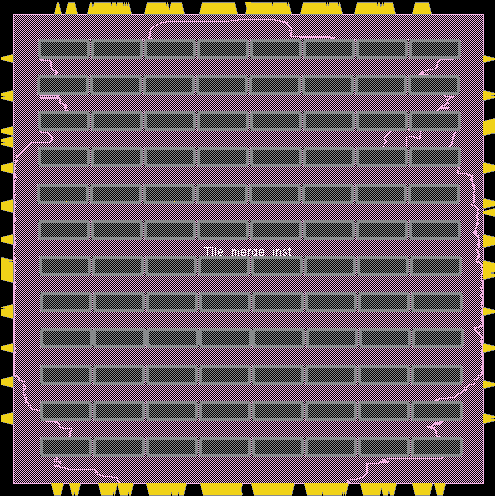}
        \label{layout}
    }

\caption{(a) The architecture of MMIE and (b) its layout.}
\end{figure*}

Eq. (\ref{cc}) suggests that the number of required clock cycles for convolutional computations is independent of $N$ for large values of $N$ (i.e., $S \times N \gg (W_f - S)$) when not considering the weight passing overheads. In Section \ref{MMIE:PT} we showed that input pixels are shared among all tiles and each pixel is read at each clock cycle. This means that the number of memory accesses by input activation maps (MA$_{imaps}$) is equal to the number of clock cycles required to complete the convolution. On the other hand, the weights are read in the first $W_f$ clock cycles out of a total $(S \times N + W_f - S)$. As a result, the number of required memory accesses by filters to compute $N$ out of $H_{out} \times W_{out}$ pixels of one out of $C_{out}$ output activation map is equal to $C_{in} \times H_f \times W_f$. In general, the number of memory accesses by filters (MA$_{filters}$) can be computed as follows:
\begin{equation}
\text{MA$_{filters}$} = H_f \times W_f \times C_{in} \times \left \lceil \dfrac{W_{out} \times H_{out}}{N} \right \rceil \times C_{out}.
\end{equation}
Finally, the total number of memory accesses (MA) is a summation of memory accesses by filters, input activation maps and output activation maps, where the number of memory accesses by output activation maps (MA$_{omaps}$) is equal to $W_{out} \times H_{out}$. It is worth noting that while the number of clock cycles and MA$_{imaps}$ are independent of $N$, MA$_{filters}$ depends on it. On the other hand, MA$_{filters}$ is independent of $p$ while the number of clock cycles and MA$_{imaps}$ are not. It is worth mentioning that while higher values of $p$ and $N$ optimize MMIE towards lower memory accesses and processing latencies, they also increases its power consumption and silicon area.
\subsubsection{Fully-Connected Computations}\label{MMIE:PTMA:fc}
In Section \ref{MMIE:WG_fc}, we showed that MMIE can perform the fully-connected computations in a similar way to convolutional computations, with each PE loading a different set of weights. The processing time of each PE is thus equal to the number of inputs $n$. The number of clock cycles required to generate $m$ output pixels can be expressed as
\begin{equation}
\text{No. CC} = \left \lceil \dfrac{m}{p} \right \rceil \times n.
\end{equation}
Unlike weights, input pixels are shared among PEs. Therefore, the number of memory accesses by input pixels (MA$_{ip}$) is equal to the number of clock cycles required for fully-connected computations. Since each output pixel relies on a distinct set of $n$ weights, the number of memory accesses by weights (MA$_{weights}$) is computed as follows:
\begin{equation}
\text{MA$_{weights}$} = m \times n.
\end{equation}
The number of memory accesses by output pixels (MA$_{op}$) is equal to $m$. The total number of memory accesses (MA) is also a summation of memory accesses by weights, input and output pixels.

\section{Implementation Results} \label{sec:HIR}
%\fixme{In Section \ref{MMIE:PTMA} we showed that the number of clock cycles required for both fully-connected and convolutional computations highly depends on $p$ while it is independent of $N$. Similarly, MA$_{imaps}$ and MA$_{ip}$ are also dependent of $p$ and independent of $N$. On the other hand, MA$_{filters}$ highly relies on the value of $N$ while MA$_{weights}$ is independent of it. It is worth mentioning that the utilization factor of PEs and consequently the performance efficiency of MMIE also relies on the value of $N$ (see Section \ref{sec:MMIE}). In fact, higher values of $N$ optimizes MMIE for achieving lower memory accesses with high performance efficiency while higher values of $p$ optimizes MMIE towards achieving a lower processing latency.}{This is a lot of repetition. We can get rid of it or summarize to save space and make it more readable.}

In this paper, we optimize MMIE for a low-latency, low-memory access implementation while keeping its power consumption below the power budget of mobile devices, limited to a few hundred mW \cite{yodann}. Fig. \ref{MMIE} shows the architecture of MMIE which is consisted of three main sub-blocks: tiles, pipelining stages and a distributor unit. MMIE contains 32 reconfigurable tiles, each of which with 6 PEs. Each PE is also associated with a $L = 64$ 24-bit memory. The pipelining stages provide the required amount of shifts depending on the value of $W_f$ using shift registers and multiplexers, as discussed in Section \ref{MMIE:PT}. The distributor unit also provides the required bandwidth for fully-connected weights using shift registers working at lower frequency than the off-chip memory.\par

The $p$ and $N$ parameters do not only affect latency and number of memory accesses, but also impact power and area costs. Therefore, it is possible to obtain different trade-offs between processing time, throughput and implementation costs depending on $p$ and $N$.
Since the reconfigurable tile functions differently based on $W_f$ and $S$, the effective values of $N$ and $p$ vary for each case. Table \ref{eff_par} shows the effective values of $N$ and $p$ for AlexNet, VGGNet and ResNet filter sizes. The effective values of $N$ and $p$, denoted as $N_{eff}$ and $p_{eff}$ respectively, have to be used in all the equations reported in this paper that rely on these two values.

MMIE was implemented in TSMC 65nm GP CMOS technology and its layout are shown in Fig. \ref{layout}. MMIE works at a nominal frequency of 200 MHz and 40 MHz for convolutional and fully-connected processes respectively. MMIE performs the fully-connected computations at a lower frequency since they require a high input bandwidth, as each neuron loads its own set of weights. We also used the run-length compression technique introduced in \cite{eyeriss1} to reduced the required bandwidth. MMIE uses the distributor unit to decode the compressed values. Considering MMIE working at 10$\times$ lower frequency compared to the off-chip memory for fully-connected computations, the required bandwidth of 193 16-bit values are obtained using this technique.

\subsection{Hardware Implementation Results on State-of-the-Art Networks}\label{HIR:breakdown}
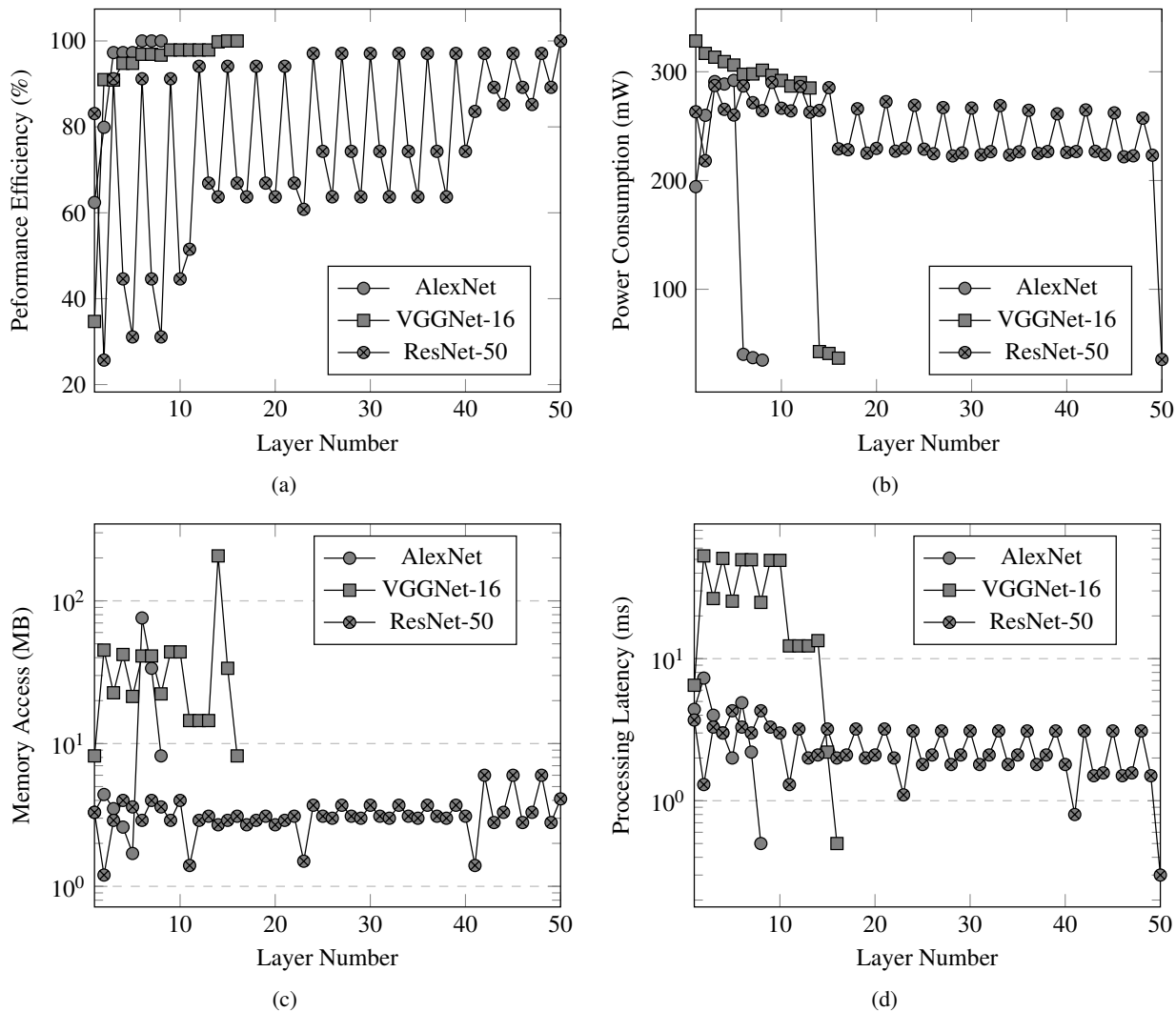
\begin{figure*}[!t]
    \centering
    
    \subfigure[]{
\small
\scalebox{1.2}{
\begin{tikzpicture}

\begin{axis}[
    title={},
    xlabel={Layer Number},
    ylabel={Peformance Efficiency (\%)},
    xmin=1, xmax=50,
    %ymin=0, ymax=8000,
    %xtick={4,8,16,32,64,128,256},
    %ytick={0,500,1000,2000,4000,8000},
    %legend pos=north east,
    legend style ={ at={(0.5,0.33)},
        anchor=north west, draw=black,
        fill=white,align=left},
    cycle list name=black white
]
\addplot
  coordinates{
    (1,62.4)
    (2,79.9)
    (3,97.3)
    (4,97.3)
    (5,97.3)
    (6,100)
    (7,100)
    (8,100)
};
\addlegendentry{AlexNet}
\addplot
  coordinates{
    (1,34.7)
    (2,91)
    (3,90.9)
    (4,94.8)
    (5,94.8)
    (6,96.9)
    (7,96.9)
    (8,96.7)
    (9,97.9)
    (10,97.9)
    (11,97.9)
    (12,97.9)
    (13,97.9)
    (14,99.8)
    (15,100)
    (16,100)
};
\addlegendentry{VGGNet-16}
\addplot
  coordinates{
    (1,83.1)
    (2,25.7)
    (3,91.2)
    (4,44.6)
    (5,31.1)
    (6,91.2)
    (7,44.6)
    (8,31.1)
    (9,91.2)
    (10,44.6)

    (11,51.5)
    (12,94.1)
    (13,66.9)
    (14,63.7)
    (15,94.1)
    (16,66.9)
    (17,63.7)
    (18,94.1)
    (19,66.9)
    (20,63.7)
    (21,94.1)
    (22,66.9)

    (23,60.8)
    (24,97.1)
    (25,74.3)
    (26,63.7)
    (27,97.1)
    (28,74.3)
    (29,63.7)
	(30,97.1)
    (31,74.3)
    (32,63.7)
    (33,97.1)
    (34,74.3)
    (35,63.7)
    (36,97.1)
    (37,74.3)
    (38,63.7)
    (39,97.1)
    (40,74.3)

    (41,83.6)
    (42,97.1)
    (43,89.2)
    (44,85.2)
    (45,97.1)
    (46,89.2)
    (47,85.2)
    (48,97.1)
    (49,89.2)
    (50,100)
};
\addlegendentry{ResNet-50}
\end{axis}
\end{tikzpicture}}
\label{PE}
}
\subfigure[]{
\small
\scalebox{1.2}{
\begin{tikzpicture}

\begin{axis}[
    title={},
    xlabel={Layer Number},
    ylabel={Power Consumption (mW)},
    xmin=1, xmax=50,
    %ymin=0, ymax=8000,
    %xtick={4,8,16,32,64,128,256},
    %ytick={0,500,1000,2000,4000,8000},
    %legend pos=north east,
    legend style ={ at={(0.5,0.33)},
        anchor=north west, draw=black,
        fill=white,align=left},
    cycle list name=black white
]
\addplot
  coordinates{
    (1,194.35)
    (2,259.96)
    (3,290.93)
    (4,288.7)
    (5,292.01)
    (6,40.2)
    (7,37.14)
    (8,34.89)
};
\addlegendentry{AlexNet}
\addplot
  coordinates{
    (1,328.3)
    (2,316.86)
    (3,313.38)
    (4,309.3)
    (5,306.14)
    (6,297.67)
    (7,297.95)
    (8,301.41)
    (9,296.81)
    (10,291.95)
    (11,286.82)
    (12,290.01)
    (13,285.05)
    (14,42.7)
    (15,41.01)
    (16,36.7)
};
\addlegendentry{VGGNet-16}
\addplot
  coordinates{
    (1,263.2)
    (2,218.17)
    (3,287.3)
    (4,265.4)
    (5,260.19)
    (6,286.9)
    (7,271.57)
    (8,264.18)
    (9,290.15)
    (10,266.44)
    (11,264.01)
    (12,286.49)
    (13,262.8)
    (14,264.42)
    (15,285.35)
    (16,229.07)
    (17,228.28)
    (18,265.96)
    (19,225.17)
    (20,229.51)
    (21,272.49)
    (22,226.95)
    (23,229.6)
    (24,269.13)
    (25,228.78)
    (26,224.61)
    (27,267.05)
    (28,222.62)
    (29,225.35)
	(30,266.59)
    (31,223.62)
    (32,226.51)
    (33,268.87)
    (34,223.43)
    (35,226.41)
    (36,264.54)
    (37,224.92)
    (38,226.67)
    (39,261.32)
    (40,225.84)
    (41,226.64)
    (42,264.88)
    (43,226.84)
    (44,223.70)
    (45,262.2)
    (46,221.8)
    (47,222.68)
    (48,257.26)
    (49,223.28)
    (50,35.5)
};
\addlegendentry{ResNet-50}
\end{axis}
\end{tikzpicture}}
\label{PC}
}
\subfigure[]{
\small
\scalebox{1.2}{
\begin{tikzpicture}

\begin{axis}[
    title={},
    xlabel={Layer Number},
    ylabel={Memory Access (MB)},
    xmin=1, xmax=50,
    %ymin=0, ymax=8000,
    %xtick={4,8,16,32,64,128,256},
    %ytick={0,500,1000,2000,4000,8000},
    %legend pos=north east,
    ymajorgrids=true,
    grid style=dashed,
    ymode = log,
    log basis y = 10,
    legend style ={ at={(0.47,0.97)},
        anchor=north west, draw=black,
        fill=white,align=left},
    cycle list name=black white
]

\addplot
  coordinates{
    (1,3.3)
    (2,4.4)
    (3,3.5)
    (4,2.6)
    (5,1.7)
    (6,75.9)
    (7,33.7)
    (8,8.2)
}; \addlegendentry{AlexNet}
\addplot
  coordinates{
    (1,8.2)
    (2,45.2)
    (3,22.7)
    (4,42.1)
    (5,21.4)
    (6,41.1)
    (7,41.1)
    (8,22.3)
    (9,43.9)
    (10,43.9)
    (11,14.5)
    (12,14.5)
    (13,14.5)
    (14,206.6)
    (15,33.7)
    (16,8.2)
};
\addlegendentry{VGGNet-16}
\addplot
  coordinates{
    (1,3.3)
    (2,1.2)
    (3,2.9)
    (4,4)
    (5,3.6)
    (6,2.9)
    (7,4)
    (8,3.6)
    (9,2.9)
    (10,4)

    (11,1.4)
    (12,2.9)
    (13,3.1)
    (14,2.7)
    (15,2.9)
    (16,3.1)
    (17,2.7)
    (18,2.9)
    (19,3.1)
    (20,2.7)
    (21,2.9)
    (22,3.1)

    (23,1.5)
    (24,3.7)
    (25,3.1)
    (26,3)
    (27,3.7)
    (28,3.1)
    (29,3)
	(30,3.7)
    (31,3.1)
    (32,3)
    (33,3.7)
    (34,3.1)
    (35,3)
    (36,3.7)
    (37,3.1)
    (38,3)
    (39,3.7)
    (40,3.1)

    (41,1.4)
    (42,6)
    (43,2.8)
    (44,3.3)
    (45,6)
    (46,2.8)
    (47,3.3)
    (48,6)
    (49,2.8)

    (50,4.1)
};
\addlegendentry{ResNet-50}
\end{axis}
\end{tikzpicture}}
\label{MA}
}
\subfigure[]{
    \small
    \scalebox{1.2}{
\begin{tikzpicture}
\begin{axis}[
    title={},
    xlabel={Layer Number},
    ylabel={Processing Latency (ms)},
    xmin=1, xmax=50,
    %ymin=0, ymax=8000,
    %xtick={4,8,16,32,64,128,256},
    %ytick={0,500,1000,2000,4000,8000},
    %legend pos=north east,
    ymajorgrids=true,
    grid style=dashed,
    ymode = log,
    log basis y = 10,
    legend style ={ at={(0.47,0.97)},
        anchor=north west, draw=black,
        fill=white,align=left},
    cycle list name=black white
]
\addplot
  coordinates{
    (1,4.4)
    (2,7.3)
    (3,4)
    (4,3)
    (5,2)
    (6,4.9)
    (7,2.2)
    (8,0.5)
};
\addlegendentry{AlexNet}
\addplot
  coordinates{
    (1,6.5)
    (2,52.9)
    (3,26.5)
    (4,50.8)
    (5,25.4)
    (6,49.7)
    (7,49.7)
    (8,24.9)
    (9,49.2)
    (10,49.2)
    (11,12.3)
    (12,12.3)
    (13,12.3)
    (14,13.4)
    (15,2.2)
    (16,0.5)
};
\addlegendentry{VGGNet-16}
\addplot
  coordinates{
    (1,3.7)
    (2,1.3)
    (3,3.3)
    (4,3)
    (5,4.3)
    (6,3.3)
    (7,3)
    (8,4.3)
    (9,3.3)
    (10,3)

    (11,1.3)
    (12,3.2)
    (13,2)
    (14,2.1)
    (15,3.2)
    (16,2)
    (17,2.1)
    (18,3.2)
    (19,2)
    (20,2.1)
    (21,3.2)
    (22,2)

    (23,1.1)
    (24,3.1)
    (25,1.8)
    (26,2.1)
    (27,3.1)
    (28,1.8)
    (29,2.1)
	(30,3.1)
    (31,1.8)
    (32,2.1)
    (33,3.1)
    (34,1.8)
    (35,2.1)
    (36,3.1)
    (37,1.8)
    (38,2.1)
    (39,3.1)
    (40,1.8)

    (41,0.8)
    (42,3.1)
    (43,1.5)
    (44,1.57)
    (45,3.1)
    (46,1.5)
    (47,1.57)
    (48,3.1)
    (49,1.5)
    (50,0.3)
};
\addlegendentry{ResNet-50}
\end{axis}
\end{tikzpicture}}
\label{PT}
}

\caption{(a) The performance efficiency, (b) power consumption, (c) memory access and (d) computation latency breakdowns of AlexNet, VGGNet-16 and ResNet-50 at 200 MHz for convolutional processes and 40 MHz for fully-connected computations in TSMC 65 nm CMOS technology.}
    \label{fig6}
\end{figure*}

Fig. \ref{PE} shows the breakdown of performance efficiency for each layer of AlexNet, VGGNet-16 and ResNet-50 when using MMIE. In our simulations, the input pixels and weight values are quantized to 16 bits while using 2 and 15 fractional bits, respectively. It is worth noting that this quantization scheme only results in less than 0.5\% accuracy degradation on the aforementioned CNNs using \cite{matconvnet,caffe}. The implementation results show that the lowest performance efficiency of AlexNet and VGGNet-16 was obtained at the first layer of these networks. The number of output activation maps $C_{out}$ of the first layer of AlexNet is 96 while MMIE provide 64 parallel tiles when $W_f = 11$ and $S = 4$. As a result, for the first 64 output activation maps, MMIE achieves a high performance efficiency while the remaining 32 output activation maps are computed using 32 parallel tiles out of 64, which explain the low performance efficiency of this layer. On the other hand, MMIE successfully performs the computations of the first layer of VGGNet-16 with a high performance efficiency. However, since the required time for writing the computed output activation pixels is longer than the computation time, the low performance efficiency is inevitable. In ResNet-50, layers with a receptive field of $1 \times 1$ show lower performance compared to other filter sizes, while it was shown in Section \ref{MMIE:WG1x1} that such receptive field yields a $100\%$ performance efficiency. Such performance efficiency degradation is expected, as $C_{out}$ of the layers with receptive field of $1 \times 1$ are not multiple of 192 available parallel tiles. For instance, the number of output activation maps of the second layer of ResNet-50 is 64, while 192 parallel tiles are available. Therefore, 128 tiles are not being used for this layer.

Fig. \ref{PC} shows the breakdown of power consumption for each layer of AlexNet, VGGNet-16 and ResNet-50. The power consumption of MMIE follows a descending trend as the number of zeros in output/input activations maps and filters increases for each layer of AlexNet, VGGNet-16 and ResNet-50. Moreover, it also increases as the performance efficiency of layers rises. The power numbers reported in this paper are obtained by measuring switching activities of all models.
\par

Fig. \ref{MA} shows the breakdown of the memory accesses for each layer of AlexNet, VGGNet-16 and ResNet-50. The memory accesses for each layer of the aforementioned networks are limited to a few MB. More precisely, AlexNet and ResNet-50 layers require a lower number of memory accesses compared to VGGNet-16. While the memory accesses for each layer of AlexNet and ResNet-50 are roughly in the same order, the total memory accesses of ResNet-50 are significantly more due to its numerous layers. The processing latency of each layer also follows a similar trend to the memory accesses as shown in Fig. \ref{PT}. In fact, the latency of each layer in AlexNet and ResNet-50 is limited to a few milliseconds while each layer of VGGNet-16 requires roughly 10$\times$ more clock cycles.

\subsection{Comparison With State-of-the-Art Implementations}\label{HIR:comp}

The implementation results of MMIE on AlexNet, VGGNet-16 and ResNet-50 are shown in Table \ref{tab_res}. As discussed in Section \ref{intro}, MCR of these networks varies depending on their sizes. Therefore, different implementation results are expected when running MMIE on these models. MMIE performs the convolutional and fully-connected computations of AlexNet within 20.8 ms and 7.6 ms while requiring 15.6 MB and 117.8 MB memory accesses to the off-chip memory, respectively. The convolutional and fully-connected processes of VGGNet-16 are performed within 421.8 ms and 16.4 ms and require 375.5 MB and 247.3 MB memory accesses, respectively. Finally, performing convolutional and fully-connected computations of ResNet-50 on MMIE requires 106.6 ms and 0.3 ms while memory accesses are 154.6 MB and 4.1 MB, respectively. Therefore, AlexNet computations require the lowest latency while its total memory accesses are roughly similar to those of ResNet-50. VGGNet-16 is the most complex network in terms of both processing latency and memory accesses. MMIE also yields 83\%, 94\% and 94\% performance efficiency for convolutional computations of AlexNet, VGGNet-16 and ResNet-50, respectively. It is worth mentioning that the performance efficiency of fully-connected computations is roughly 100\% for all the aforementioned networks.\par

\begin{table*}[!t]
\renewcommand{\arraystretch}{1.3}
\renewcommand{\thefootnote}{\alph{footnote}}
\caption{Comparison of the Baseline Architecture with State-of-the-art Implementations.} % title of Table
\centering % used for centering table
\scalebox{0.6}{
\large
\begin{threeparttable}

\centering
\begin{tabular}{c | c c c c c c c c c c} % centered columns (4 columns)
\hline\hline %inserts double horizontal lines
\multicolumn{1}{l}{Reference} & ISSCC'17 \cite{DNPU} & \multicolumn{2}{c}{ISSCC'17 \cite{moons_new}}
   &  \multicolumn{2}{c}{JSSC'17 \cite{eyeriss1}} & TCAS'17 \cite{TCAS} &  \multicolumn{3}{c}{This work}\\
\hline
\multicolumn{1}{l}{Technology} & NA/65 nm & \multicolumn{2}{c}{UTBB FD-SOI/28 nm} &  \multicolumn{2}{c}{TSMC/65 nm} &  TSMC/65 nm  &  \multicolumn{3}{c}{TSMC/65 nm} \\
%\multicolumn{1}{l}{Methodology} & Silicon & \multicolumn{2}{c}{Silicon} &  \multicolumn{2}{c}{Silicon} &  Synthesis &  \multicolumn{3}{c}{Layout} \\
\multicolumn{1}{l}{Gate Count\tnote{*} (NAND-2)}  & NA & \multicolumn{2}{c}{1950k} & \multicolumn{2}{c}{1852k} & 1117k  &  \multicolumn{3}{c}{1036k}\\
\multicolumn{1}{l}{Core Area\tnote{*} (mm$^2$)} & 16 & \multicolumn{2}{c}{1.87} & \multicolumn{2}{c}{12.52} &  3.5  &  \multicolumn{3}{c}{6 (2.45$\times$2.45)}\\
\multicolumn{1}{l}{\# PE} & 768(16b)-3072(4b)\tnote{c}, 64\tnote{f} & \multicolumn{2}{c}{256(16b)-1024(4b)} & \multicolumn{2}{c}{168} & 192 &  \multicolumn{3}{c}{192\tnote{c,f}}\\
\multicolumn{1}{l}{On-chip SRAM (kB)} & 290 & \multicolumn{2}{c}{144} & \multicolumn{2}{c}{181.5} & 86  &  \multicolumn{3}{c}{36.9}\\
\multicolumn{1}{l}{Nominal Frequency (MHz)} & 50-200\tnote{c,f} & \multicolumn{2}{c}{200\tnote{c}} & \multicolumn{2}{c}{250\tnote{c}} & 200\tnote{c}  &  \multicolumn{3}{c}{200\tnote{c}, 40\tnote{f}}\\
\multicolumn{1}{l}{Peak Performance (Gops)} & 300(16b)-1200(4b)\tnote{c}, 25\tnote{f} & \multicolumn{2}{c}{102(16b)-408(4b)\tnote{c}} & \multicolumn{2}{c}{84\tnote{c}} & 76\tnote{c}  &  \multicolumn{3}{c}{76.8\tnote{c}, 15.4\tnote{f}}\\
\multicolumn{1}{l}{Bitwidth (bits)} & 4-16 programmable\tnote{c}, 4-7\tnote{f} & \multicolumn{2}{c}{1-16 programmable\tnote{c}} & \multicolumn{2}{c}{16 fixed\tnote{c}} & 16 fixed\tnote{c}  &  \multicolumn{3}{c}{16 fixed\tnote{c,f}}\\
\hline
\multicolumn{1}{l}{CNN type for ImageNet}      & AlexNet & AlexNet   & VGG-16  & AlexNet   & VGG-16    & VGG-16  &  AlexNet & VGG-16 & ResNet-50\\
\multicolumn{1}{l}{Top-1 Error (\%)}              & 42.9 & 42.9  & 27  & 42.9  & 27  & 27 & 42.9 & 27 & 20\\
\multicolumn{1}{l}{Voltage (V)}                  & 0.77-1.1 & NA & NA    & 1           & 1           & 1 & 1 & 1 & 1 \\
\multicolumn{1}{l}{Power\tnote{*} (mW)}          & 63\tnote{c}, 3.5\tnote{f} & 44\tnote{c} &26\tnote{c}  & 278\tnote{c}  & 236\tnote{c}  &  260\tnote{c}    & 265\tnote{c}, 37\tnote{f}  & 301\tnote{c}, 40\tnote{f} & 248\tnote{c}, 35.5\tnote{f}\\
\multicolumn{1}{l}{Total Latency (ms)}           &  5.7\tnote{c}, 0.8\tnote{f} & 21.3\tnote{c}  &  598.8\tnote{c} & 115.3\tnote{c}   & 4309.5\tnote{c}    &  453.3\tnote{c} & 20.8\tnote{c}, 7.6\tnote{f} & 421.8\tnote{c}, 16.4\tnote{f} & 103.6\tnote{c}, 0.3\tnote{f}\\
\multicolumn{1}{l}{Throughput (fps)}             & 177\tnote{c}, 1.2k\tnote{f} & 47\tnote{c} & 1.67\tnote{c}  & 34.7\tnote{c}  & 0.7\tnote{c}  &  2.21\tnote{c} & 48.1\tnote{c}, 131.6\tnote{f} & 2.2\tnote{c}, 61\tnote{f} & 9.6\tnote{c}, 3.3k\tnote{f}\\
\multicolumn{1}{l}{Performance (Gops)}           & 235.4\tnote{c}, 140.6\tnote{f} & 62.6\tnote{c} & 51.3\tnote{c} & 46.1\tnote{c}  & 21.4\tnote{c}  &  67.7\tnote{c} & 63.9\tnote{c}, 15.4\tnote{f} & 72.5\tnote{c}, 15.1\tnote{f} & 74.5\tnote{c}, 15\tnote{f}\\
\multicolumn{1}{l}{Performance Efficiency}       & 50\%\tnote{c}, 562\%\tnote{f} & 38\%\tnote{c} & 32\%\tnote{c} & 55\%\tnote{c} & 26\%\tnote{c} & 89\%\tnote{c} & 83\%\tnote{c}, 100\%\tnote{f} & 94\%\tnote{c}, 98\%\tnote{f} & 88\%\tnote{c}, 97\%\tnote{f}\\
\multicolumn{1}{l}{Energy-Efficiency\tnote{*} (Gops/W)}  & 4200\tnote{c}, 40.2k\tnote{f} & 1423\tnote{c}  & 1973\tnote{c} & 166\tnote{c}  & 90.7\tnote{c}  & 260.4\tnote{c}    & 241.1\tnote{c}, 416.2\tnote{f}      & 240.9\tnote{c}, 377.5\tnote{f}        & 300.4\tnote{c}, 422.5\tnote{f}\\
\multicolumn{1}{l}{Memory Access / Batch (MB)}   & NA & NA   &  NA   & 15.4\tnote{c}        & 321.1\tnote{c}       & 331.7\tnote{c}  & 15.6\tnote{c}, 117.8\tnote{f} & 375.5\tnote{c}, 247.3\tnote{f} & 154.6\tnote{c}, 4.1\tnote{f}\\
\hline
\hline
\end{tabular}
\begin{tablenotes}
      \normalsize
      \item[*] Including on-chip SRAM.
      \item[f] Fully-connected.
      \item[c] Convolutional.
\end{tablenotes}
\end{threeparttable}}

\label{tab_res} % is used to refer this table in the text
\vspace{-10pt}
\end{table*}

During the past few years, numerous works have been conducted towards ASIC implementations of DNNs. However, most of them were only tested on either small datasets or outdated CNNs which require order of magnitudes lower parameters and computations \cite{dac,origami2,origami3,yodann,conv_imp1}. Recently, Google released a custom DNN accelerator tensor processing unit (TPU) \cite{TPU}. TPU is a programmable and reconfigurable accelerator that can perform both fully-connected and convolutional computations. However, its power consumption exceeds the power budgets of embedded devices \cite{TCAS}. In \cite{eyeriss1,eyeriss}, a convolutional accelerator, called Eyeriss, was introduced. Eyeriss was fabricated in 65 nm CMOS technology and tested on AlexNet and VGGNet-16. Eyeriss uses high batch sizes to obtain a lower number of memory accesses, but using this method results in a higher computational latency. Eyeriss performs convolutional computations of AlexNet and VGGNet-16 in 115.3 ms and 4.3 s while requiring 15.4 MB and 321.1 MB memory accesses and using batch size of 4 and 3, respectively. Its performance efficiency is also limited to only 55\% and 26\% on AlexNet and VGGNet-16, resulting in large silicon area of 12.52 mm$^2$ (1852kgates). Eyeriss also uses clock gating to reduce its power consumption. \par

Recently, a few works have focused on minimizing energy by modulating precision, frequency and supply voltage of their accelerator for each convolutional layer \cite{moons1,moons2, moons_new}. In \cite{moons_new}, a precision-scalable convolutional accelerator, fabricated in 28 nm UTBB FD-SOI technology, was introduced. This architecture dynamically adapts itself depending on the required precision for each layer, instead of using a fixed precision. More precisely, it exploits a reconfigurable multiplier which is able to perform a 16-bit, two 8-bit and four 4-bit multiplications, depending on the required precision. As a result, using a dynamic fixed-point technique allows to change frequency and supply voltage over time which results in a lower power/energy consumption. This accelerator performs the convolutional computations of AlexNet to 21.3 ms, and those of ResNet to 598.8 ms, while its performance efficiency is respectively limited to 38\% and 32\% on average. Similar to Eyeriss, the low performance efficiency of this architecture results in a large gate count of 1950kgates.\par

In \cite{DNPU}, a DNN accelerator, fabricated in 65 nm CMOS 1P8M, was introduced. This accelerator can perform both fully-connected and convolutional computations while using two separate cores and the dynamic fixed-point technique to minimize power/energy consumption. This architecture exploits a reconfigurable 16-bit multiplier for convolutional processes which allows it to work with lower frequency and supply voltage. This architecture performs convolutional and fully-connected computations within 5.7 ms and 833 $\mu$s, respectively. The convolutional core of this architecture contains 768 16-bit reconfigurable PEs, which can be used as 3072 4-bit PEs. Despite its high convolutional throughput, its performance efficiency is limited to $50\%$ on average. The fully-connected core contains only 64 PEs, and uses a quantization table-based matrix multiplication to reduce off-chip memory accesses and remove redundancy. This technique reduces the memory accesses by $75\%$ and avoids $90\%$ of the 16-bit fixed-point multiplications in fully-connected computations \cite{DNPU}. While the fully-connected core is highly optimized, it requires separate PEs and hardware resources, which leads to a large silicon area of 16 mm$^2$.\par

In \cite{TCAS}, a convolutional accelerator was proposed as a first attempt to improve the performance efficiency for filters fixed to $3 \times 3$. This architecture performs the convolutional computations of VGGNet-16 within 453.3 ms and requires 331.7 MB memory accesses.

In this paper, we proposed MMIE which supports all the filter sizes that require less than or equal to 6 parallel PEs in each tile. MMIE can perform both the convolutional and fully-connected computations while using the same PEs. Since both Eyeriss and MMIE were implemented in TSMC 65nm CMOS technology and use 16-bit fixed-point representations, a direct comparison of these two implementations constitutes a fair comparison. As shown in Table \ref{tab_res}, MMIE outperforms Eyeriss \cite{eyeriss1} in terms of gate count (1.8$\times$ smaller), latency (5.5$\times$ and 10.2$\times$ lower), throughput (1.4$\times$ and 3.1$\times$ faster), performance efficiency (1.5$\times$ and 3.6$\times$ better) and energy scalability (1.5$\times$ and 2.7$\times$ more efficient) while having roughly the same number of memory accesses per batch. It is worth noting that a direct comparison of MMIE with the works published in \cite{moons_new,DNPU} does not constitute a fair comparison, since they dynamically modulate precision, frequency and supply voltage and use advanced technology nodes, which allows them to instantiate more PEs while still having a low-power/energy consumption. However, the introduced performance efficiency metric can be used for a fair comparison as it reflects the performance of the accelerators independent of their technology nodes, precisions and optimization techniques. Therefore, MMIE has better the performance efficiency than the works published in \cite{DNPU} (2$\times$ better) and \cite{moons_new} (2.2$\times$ and 2.9$\times$ better) when performing convolutions of AlexNet and VGGNet-16.

\section{Conclusion}\label{conclusion}
CNN accelerators in literature promise a high peak throughput, but their performance is limited to less than 55 \% when running the state-of-the-art networks such as AlexNet, VGGNets and ResNets. We proposed a dataflow inspired to the fully-connected computations to perform both convolutional and fully-connected processes with a high utilization factor. We then introduced a multi-mode inference engine (MMIE) based on the proposed dataflow and theoretically formalized its implementation performance. Finally, we implemented MMIE in TSMC 65nm CMOS technology and tested it on three state-of-the-art networks, AlexNet, VGGNet-16 and ResNet-50. The implementation results show that MMIE performs both the fully-connected and convolutional computations with performance efficiency no less than 84\%, outperforming the state of the art also in terms of area occupation.

%%%%%%% -- PAPER CONTENT ENDS -- %%%%%%%%

%%%%%%%%% -- BIB STYLE AND FILE -- %%%%%%%%
\bibliographystyle{ieeetr}
\bibliography{Bibliography}
%%%%%%%%%%%%%%%%%%%%%%%%%%%%%%%%%%%%

\end{document}